\pdfoutput=1
\documentclass[a4paper,12pt]{article}

\usepackage{amsfonts}
\usepackage{mathrsfs}
\usepackage{amsmath}
\usepackage{amssymb}
\usepackage{framed} 
\usepackage[medium]{titlesec}
\usepackage{bm}
\usepackage{cite} 
\usepackage[normalem]{ulem}
\usepackage{extarrows}
\usepackage{slashed}
\usepackage{isodateo}
\usepackage{graphicx}
\usepackage{xcolor}
\usepackage[bookmarksnumbered=true,bookmarksopen=true]{hyperref}
 \hypersetup{colorlinks,%
             linkcolor=[rgb]{0,0.3,0.6}, %
             citecolor=[rgb]{0,0.3,0.6}, %
             urlcolor=[rgb]{0,0.3,0.6}}
\usepackage[hmargin=.7in,vmargin=1.1in]{geometry}
\usepackage{indentfirst}
\newcommand{\FR}[2]{\displaystyle\frac{\,{#1}\,}{#2}}
\newcommand{\fr}[2]{\mbox{$\frac{\,{#1}\,}{#2}$}}
\newcommand{\n}{\nonumber}
\renewcommand{\rm}{\mathrm}

\graphicspath{{fig/}}

\def\bge{\begin{equation}}
\def\ede{\end{equation}}
\def\bga{\begin{aligned}}
\def\eda{\end{aligned}}
\def\bgp{\begin{pmatrix}}
\def\edp{\end{pmatrix}}
\def\bgs{\begin{subequations}}
\def\eds{\end{subequations}}
\newcommand{\order}[1]{\mathcal{O}({#1})}
\def\di{{\mathrm{d}}}
\def\D{{\mathrm{D}}}

\def\mb{\mathbf}

\def\pd{\partial}
\def\ld{{\mathscr{L}}}

\def\la{\langle}\def\ra{\rangle}

\def\to{\rightarrow}

\def\ii{\mathrm{i}}

\def\al{\alpha}
\def\be{\beta}
\def\ga{\gamma}
\def\de{\delta}
\def\ep{\epsilon}
\def\ka{\kappa}
\def\lam{\lambda}
\def\rh{\rho}
\def\si{\sigma}

\def\Mp{M_{\text{Pl}}}

\newcommand{\ob}[1]{\mkern 2mu \overline{\mkern -2mu #1 \mkern -2mu}\mkern 2mu}
\newcommand{\wt}[1]{\mkern 2mu \widetilde{\mkern -2mu #1 \mkern -2mu}\mkern 2mu}

\begin{document}  

\title{\Large\textbf{Gauge Boson Signals at the Cosmological Collider}} 
\author{Lian-Tao Wang$^a$\footnote{Email: liantaow@uchicago.edu}~~~and~~~Zhong-Zhi Xianyu$^b$\footnote{Email: zxianyu@g.harvard.edu}\\[2mm]
\normalsize{\emph{$^a$~Department of Physics, University of Chicago, Chicago, IL 60637}}\\
\normalsize{\emph{$^b$~Department of Physics, Harvard University, 17 Oxford Street, Cambridge, MA 02138}}}

\date{}

\maketitle
 
\begin{abstract}
  We study the production of massive gauge bosons during inflation from the axion-type coupling to the inflaton and the corresponding oscillatory features in the primordial non-Gaussianity. In a window in which both the gauge boson mass and the chemical potential are large, the signal is potentially reachable by near-future large scale structure probes. This scenario covers a new region in oscillation frequency which is not populated by previously known cosmological collider models. We also demonstrate how to properly include the exponential factor and discuss the  subtleties in obtaining power dependence of the gauge boson mass in the signal estimate.  
\end{abstract}

\section{Introduction}

Cosmological inflation in the early universe sets the stage for rich dynamics of particle physics at  energy scales much above the reach of terrestrial experiments. In the coming decades, much more observational data will further shed light in this era. In particular, the precision in the primordial Non-Gaussianity (NG) measurement will be improved by orders of magnitudes \cite{Meerburg:2019qqi}. Among various NG observables, the oscillatory shape in the squeezed limit due to particle production during the inflation is particularly striking. (We will henceforth refer to this oscillatory shape the ``signal.'') Detecting such a signal at this so-called cosmological collider offers direct evidence of new physics particles and a tool of studying their properties \cite{Chen:2012ge,Chen:2009zp,Arkani-Hamed:2015bza,Chen:2016nrs,Lee:2016vti,Meerburg:2016zdz,Chen:2016uwp,Chen:2016hrz,An:2017hlx,Kumar:2017ecc,Chen:2018xck,Wu:2018lmx,Li:2019ves,Lu:2019tjj,Hook:2019zxa,Hook:2019vcn,Kumar:2019ebj,Wang:2019gbi,Wang:2020uic,Li:2020xwr}.

\begin{figure}[h!]
\centering
\includegraphics[width=0.55\textwidth]{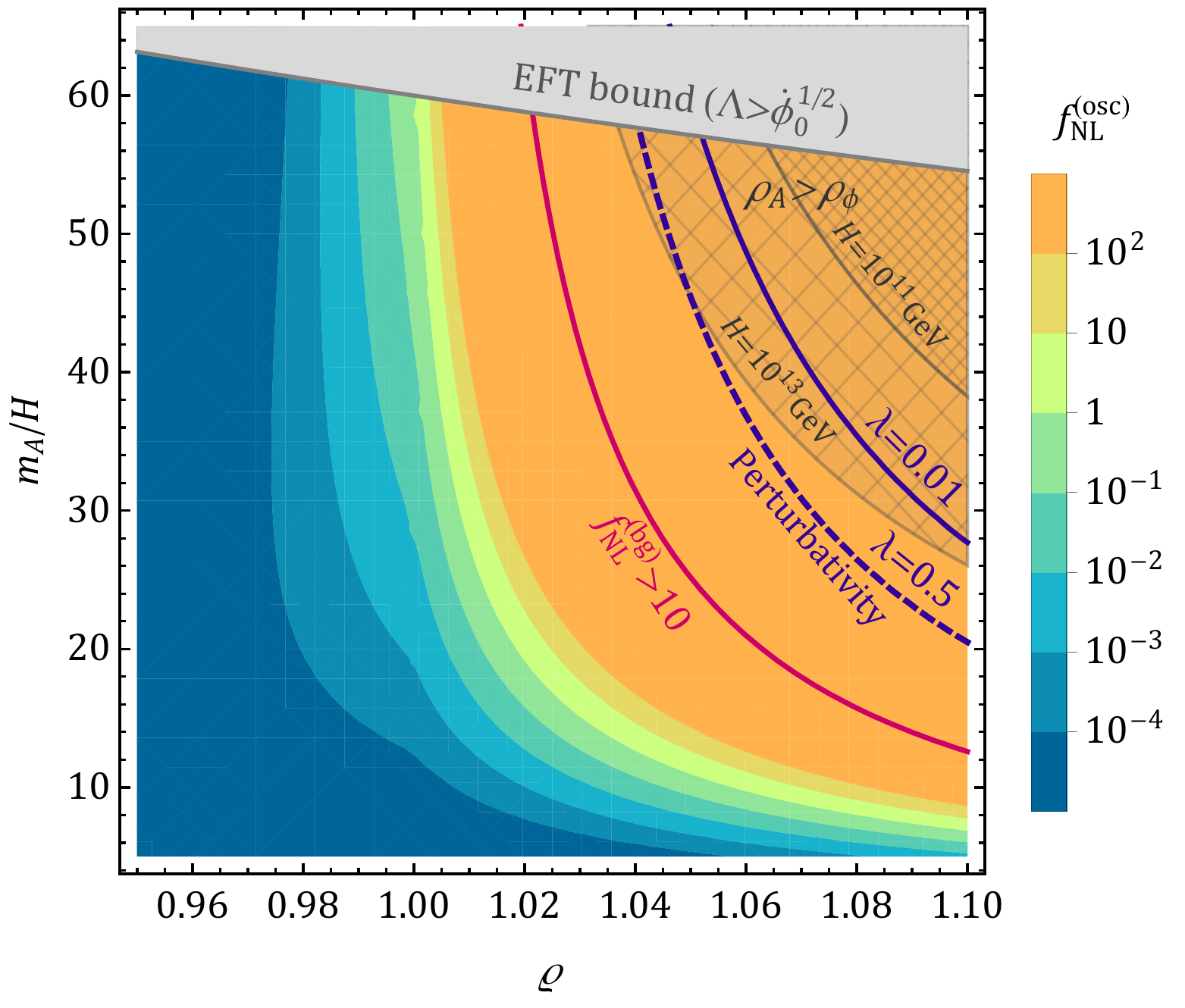}
\caption{The signal size of the gauge-boson-mediated oscillatory NG as a function of chemical-potential-mass-ratio $\varrho=\mu/m_A$ and the gauge boson mass $m_A$ in the unit of Hubble $H$ [Eqs. (\ref{fnl_b_simp}) and (\ref{fnl_c_simp})]. In this plot we take $u=1$ which is defined in (\ref{par}). Also shown in the figure are the constraints from the validity of EFT expansion (upper grey region), no large back reaction to inflation dynamics (meshed regions), the validity of perturbative calculation (solid and dashed blue curves), and the associated equilateral NG (magenta curve). See Sec.\;\ref{sec_constraints} for more discussions.}
\label{fig_fnl}
\end{figure}

\begin{figure}[h!]
\centering
\includegraphics[width=0.55\textwidth]{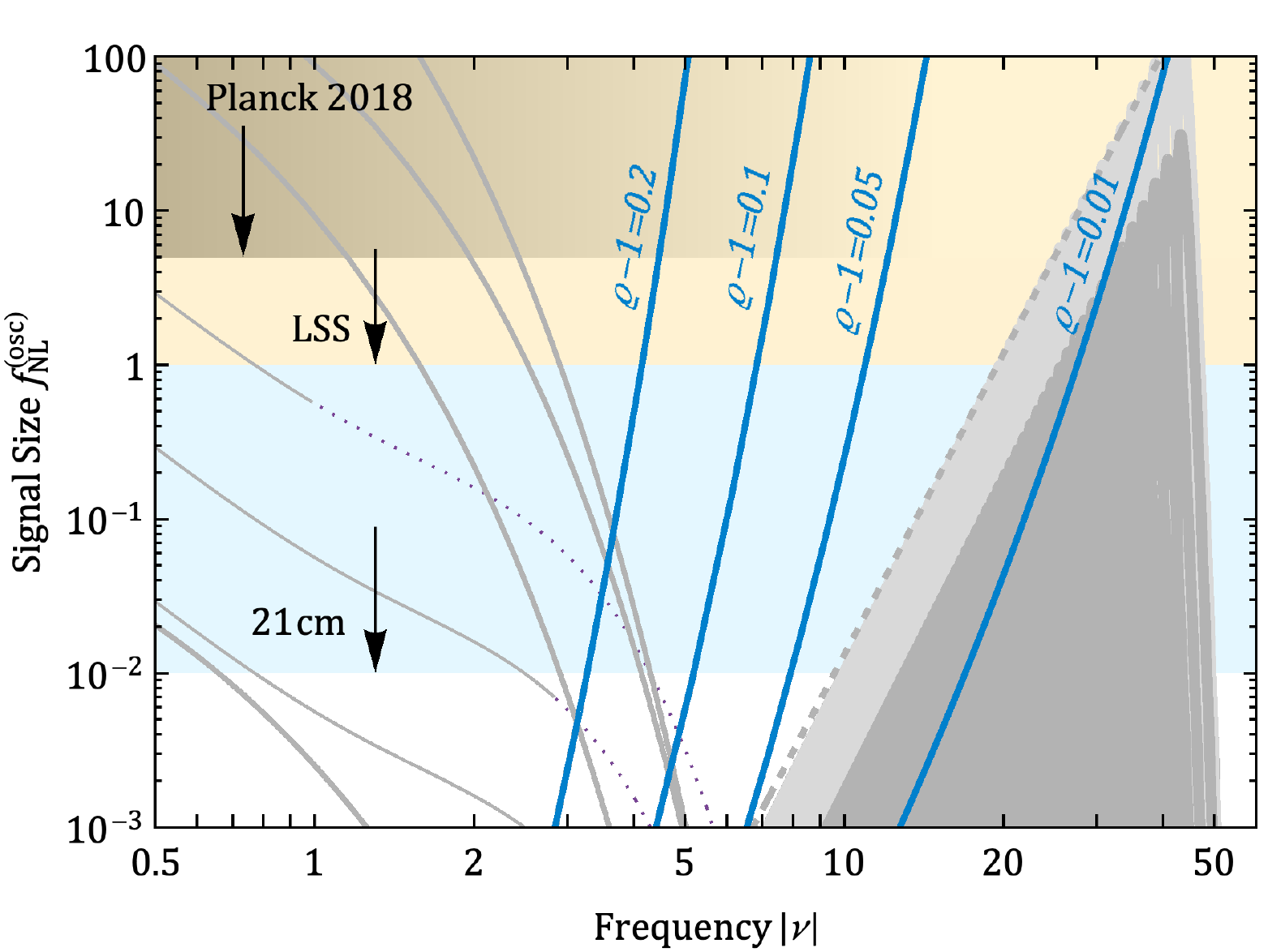}
\caption{The same signals as in Fig.\;\ref{fig_fnl} plotted as functions of oscillation frequency $\wt\nu=\sqrt{(m_A/H)^2-1/4}$ for several choices of $\varrho\simeq \mu/m_A$ (blue curves), together with signals from other scenarios adapted from Fig.\;1 in \cite{Wang:2019gbi}, and rough reaches of current and future observations. The light gray contours are the predictions of the quasi-single-field inflation scenario and several other scenarios, and the light gray shaded region on the right is the signal of the models in which the inflaton has a coupling to the chiral current of massive fermions.  Note that the gauge boson signals could show up in a parameter region ($4\lesssim \wt\nu\lesssim20$, $f_\text{NL}^\text{(osc)}\gtrsim 0.1$) not populated by previously known scenarios.}
\label{fig_signal}
\end{figure}

The strength of the signal depend sensitively on the coupling between the inflaton and the new physics particles. One key difference between the cosmological collider and a terrestrial collider experiment is that the interaction with the inflaton can change the spectrum of the new physics particles significantly. Very often, the signal size is exponentially sensitive to the mass. Hence, we will only have observable signals with couplings of specific types \cite{Wang:2019gbi}. This leads us to focus on a specific class of couplings by assuming the inflaton has an approximate shift symmetry, $\phi \to \phi+c$, which is well motivated by the requirement of slow roll inflation. In \cite{Wang:2019gbi}, it is further argued that a sub-class of such couplings are particularly promising. They are
\begin{equation}
\label{couplings}
\frac{1}{\Lambda}\partial_\mu \phi J^{\mu5}, \ \quad \frac{1}{\Lambda} \phi F\wt F,
\end{equation}
where $J^{\mu5}$ is a chiral fermion current and $F$ is the field strength of a gauge field. The cut off $\Lambda$ parameterizes the scale of the physics which is responsible for the generation of these operators.  These couplings introduce new sources of particle production during inflation and thus additional enhancement to the signal. In the case of the axial coupling to a fermion of mass $m$, the chemical potential $\mu=\dot\phi_0/\Lambda$ introduces both an enhancement $e^{\pi\mu/H}$ (through the particle production) and the Boltzmann suppression $e^{-\pi\sqrt{m^2+\mu^2}/H}$ (through the mass correction). Therefore, when $\mu\gg m$, the enhancement can help to largely cancel the Boltzmann suppression, leaving an $\order{1}$ rate for particle production. Using this mechanism to generate large signals has been explored in several directions in the context of cosmological collider physics \cite{Chen:2018xck,Hook:2019zxa,Hook:2019vcn}. 

The case of gauge boson production is more subtle. In the  case where the gauge boson has mass of $\order{H}$, the coupling $\phi F\wt F/\Lambda$ does not work by itself to generate the signal. On the one hand, a low cutoff $\Lambda$ would lead to exponentially fast and thus potentially dangerous particle production, which scales as power of $e^{\pi(\mu-m_A)/H}$. On the other hand, to avoid the exponential factor $e^{\pi(\mu-m_A)/H}$, we would want $\mu=\dot\phi/\Lambda\sim H$ which implies a relatively high cutoff $\Lambda\sim\dot\phi_0/H\sim 3\times 10^3H$. This will suppress the signal size which scales as $(H/\Lambda)^3$. More careful estimate shows that there is no viable intermediate range by varying the cutoff scale alone.

In this paper, we would like to emphasize that
the scenario where $\mu\sim m_A\gg H$ is much more interesting. First of all, this is a plausible scenario.  We in general expect that the heavy gauge boson gets its mass $m_A=g \si_0$ from the background value of some Higgs field $\si_0$. It is more natural to consider that $\sigma_0$ is linked to dynamics of the inflation.  In this case, we expect it to be higher than $H$, which is parametrically lower than the inflation scale. For the size of the signal,  $\mu \gg H$ allows us to have a lower cutoff scale $\Lambda$,  as long as the EFT bound $\Lambda\gtrsim \dot\phi_0^{1/2}$ is satisfied. This will help to relieve the coupling suppression without introducing the large exponential factor. Moreover, the Higgs can mix with the inflaton. Therefore, we have additional graphs generating the signal through the Higgs-gauge coupling and the Higgs-inflaton mixing, which is in general much larger than the graph with the axion coupling alone. 

Studying the above signals more carefully is the main purpose of this paper. The main results of this paper are summarized in Figs.~\ref{fig_fnl} and \ref{fig_signal}. In Fig.~\ref{fig_fnl} we show the signal size of the oscillatory features as functions of gauge boson mass $m_A$ and the ratio $\varrho\equiv \mu/m_A$, together with several constraints. The main conclusion is that there are parameter regions with sizable signals and consistent with current constraints. In Fig.\;\ref{fig_signal} we contrast this signal with signals from previously studied models, together with the expected observation reach. The main message here is that the gauge boson signals studied here could populate an intermediate mass range $5\lesssim m/H\lesssim 20$ with sizable signal size, which is in general not possible for other channels.

In  \cite{Wang:2019gbi}, we developed a simple way of estimating the sizes of the signal based largely on power counting of the relevant couplings, loop factors, and propagators.  For the purpose of studying the signals of gauge boson production, we refined such estimates to include appropriate exponential factors in this paper. We also performed a more detailed study of the behavior of the gauge boson propagator.  The main takeaways are 1) It is easy to estimate the parameter dependence correctly for the ``EFT part'' of the NG, and also easy to estimate correctly the exponential factor of the ``signal.'' However, it is in general hard to get the power dependence in the ``signal part.'' 2) The oscillatory signals are contributed by both the ``non-local'' and ``local'' part of the propagator in the late time limit, which was overlooked by many previous studies.

Gauge boson productions through an axionic coupling have been studied in different contexts \cite{Anber:2006xt, Durrer:2010mq,Barnaby:2010vf,Peloso:2016gqs,Bugaev:2013fya,Anber:2009ua,Anber:2012du}. In contrast to our paper, most of these studies focused on the scenario in which the gauge boson is massless. Ref.~\cite{Adshead:2016omu} considered the possibility that the gauge symmetry is Higgsed, to achieve better consistency with the CMB power spectrum and gravity wave measurements. In this paper, we will  focus on the production of an Abelian gauge boson. There could also be production of non-Abelian gauge bosons which could has additional interesting consequences for the cosmological collider physics. The presence of gauge field could also source a large tensor mode in the primordial fluctuation \cite{Maleknejad:2011jw, Adshead:2012kp,Adshead:2013qp}.

The rest of this paper is organized as follows. In Sec.\;\ref{sec_frame} we provide model motivations for the gauge boson signal. We figure out the signal in Sec.\;\ref{sec_signal} and various constraints in Sec.\;\ref{sec_constraints}. We conclude in Sec.\;\ref{sec_discussions}. 
In App.\;\ref{app_mass_dep} we collect some discussions about signal estimate and also about the late-time expansion of the propagators.

\section{Framework}
\label{sec_frame}

In this section, we layout the framework for our analysis. 
The starting point is an inflaton endowed with an approximate shift symmetry $\phi \to \phi + c$. An extensively studied class of inflation models with such a shift symmetry is the axion inflation, which  has a long history with many possible scenarios \cite{Freese:1990rb,Silverstein:2008sg,McAllister:2008hb,Kim:2004rp,Berg:2009tg,Dimopoulos:2005ac,Adshead:2012kp} (see \cite{Pajer:2013fsa} for a review).  Motivated by this, we consider the scenario with the following Lagrangian,
\begin{align}
\label{lag}
  \ld=&-\sqrt{-g}\bigg[\FR{1}{2}(\pd_\mu\phi)^2+V(\phi)+|\D_\mu\Sigma|^2-m_\Sigma^2 |\Sigma|^2+ \lam|\Sigma|^4\n\\
  &+\FR{1}{\Lambda_{\Sigma}^2}(\pd_\mu\phi)^2|\Sigma|^2+\FR{1}{4}F_{\mu\nu}F^{\mu\nu}\bigg]-\FR{1}{4\Lambda_{F}}\phi F_{\mu\nu}\wt F^{\mu\nu}.
\end{align}
Here $\phi$ is the inflaton and $\Sigma$ is a complex scalar charged under a local $U(1)$, $\D_\mu\Sigma=(\pd_\mu+\ii g A_\mu)\Sigma$. The rolling of the inflaton field $\la\pd_\mu\phi\ra=\dot\phi_0\de_{\mu0}$ then generates a vacuum expectation value (VEV) for $\Sigma$. Using the parameterization $\Sigma=(\si+\ii\pi)/\sqrt{2}$, 
we can write $\la\si\ra^2=\si_{0}^2=(\dot\phi_0^2/\Lambda_\Sigma^2+m_\Sigma^2)/\lam$. Here $m_\Sigma^2$ can have either sign, but we assume that the combination $\dot\phi_0^2/\Lambda_\Sigma^2+m_\Sigma^2$ is positive so that $\si$ picks up nonzero VEV. Then the scalar mass in this minimum is $m_\si^2 = 2\lam\si_{0}^2$, while the gauge boson mass is $m_A=g\si_{0}$. In addition, in the rolling inflaton background, we have a chemical potential to the gauge boson, $\mu=\dot\phi_0/\Lambda_F$. 

In general,  $m_A$, $m_\si$, and $\mu$ are free parameters in this scenario.  We will be focusing on the case $\mu \sim m_A$ in this paper. To represent the relevant parameter space, we begin with an special limit in which $m_{\Sigma}^2 \ll \dot \phi_0^2/\Lambda_\Sigma^2$.  Hence, we have the VEV $\ob \si_0^2=\dot\phi_0^2/(\lam\Lambda_\Sigma^2)$, and $\ob m_A=g\ob \si_0=g\dot\phi_0/(\sqrt{\lam}\Lambda_\Sigma)$.  If we consider $g\sim\lam\sim\order{1}$,   both masses $m_A$ and $m_\si$ will be much higher than the Hubble if the cutoff scale $\Lambda_\Sigma$ is close to its lower bound $\dot\phi_0^{1/2}$.  If we further assume $\Lambda_\Sigma = \Lambda_F $, the chemical potential $\mu=\dot\phi_0/\Lambda_F$ can be close to $m_A$. Of course, we will consider more general cases beyond this simple limit. To this end, we introduce a new parameter
\begin{align}
\label{par}
&u\equiv\FR{ \si_{0}^2}{\ob \si_{0}^2}=1+\FR{m_\Sigma^2\Lambda_\Sigma^2}{\dot\phi_0^2}. 
\end{align} 
$u -1$ measures the deviation from the limit $m_\Sigma=0$. In general, we could have any $u>0$. However, we expect $u\ll 1$ to be tuned. In general, the cut offs $\Lambda_\Sigma$ and $\Lambda_F$ could be different. We take this into account by trade $\Lambda_F$ with the chemical potential and treat it as a free parameter. In practice, it is more convenient to define the ratio $\varrho \equiv \mu/m_A$ and use it instead of $\mu$.

Next, we discuss possible models relevant for the signal we study in this paper.  Our discussion here is not  aiming at constructing a specific model. Instead, it is to motivate the corresponding parameter space of interest through some general consideration and examples. In addition to those mentioned here, there could certainly be other scenarios which can give rise to similar signals.

We begin with the energy scales involved in the problem.  
The scale of the inflation, $\Lambda_{\rm{inf}}\equiv \rho_\text{inf}^{1/4}$, is constrained by the current observation to be at most $\sim 10^{16} $ GeV. For the following discussion, we will take this upper bound as a benchmark  value. This in turn sets the Hubble scale to be $H = \Lambda_{\rm{inf}}^2/\sqrt{3}\Mp \sim 10^{13}$ GeV. The chemical potential is set by a dimension-5 operator in the inflation background, $\mu = \dot\phi_0/\Lambda_F$. From the validity of the EFT expansion, we have $\Lambda_F \geq \dot \phi_0^{1/2}$. Hence, $\mu < \dot \phi_0^{1/2} \simeq 60 H$. For reasons we will discuss in detail, we would focus on the case in which $\mu \sim m_A$. Therefore, we would be mostly interested in considering $\mu \sim m_A \sim 10^{14 \sim 15}$ GeV. 

 Since the inflaton has a shift symmetry,
it is natural to consider it as a pseudo-Goldstone boson resulting from some spontaneous symmetry breaking, with the scale $f$ which is also called the decay constant of the inflaton. At the same time, as in many familiar examples, such a Goldstone could be non-linearly realizing a symmetry which has anomalies, resulting in a coupling of the form
\begin{align}
\frac{1}{16 \pi^2} \frac{\phi}{f_{\rm{eff}}} F\wt F.
\label{eq:axionphoton}
\end{align}
$f_{\rm{eff}}$ is an effective decay constant characterizing the specific coupling between the inlfaton $\phi$ and the gauge field $F$. This $f_\text{eff}$ is to be distinguished from the scale of the symmetry breaking, $f$.  

While there is a general expectation $f \leq \Mp$ \cite{Banks:2003sx,ArkaniHamed:2006dz,Rudelius:2015xta,Heidenreich:2015wga,Heidenreich:2019bjd}, the decay constant of a generic Goldstone boson can be a free parameter otherwise. 
However, specializing to the case of the inflaton, it is more natural to consider that $f$ is related to the dynamics around the scale of inflation, and hence not too far away from $\Lambda_{\rm{inf}}$ (possibly deferring by small couplings and loop factors). For example, the models from string compactifications considered in \cite{Flauger:2009ab} have a range $f/\Mp > 10^{-4}$. 
Hence, a typical benchmark would have $f \sim 10^{16}$ -- $10^{17}$ GeV. From Eq.~(\ref{lag}) and Eq.~(\ref{eq:axionphoton}), we have $\mu \sim \dot \phi_0/(16 \pi^2 f_{\rm{eff}})$. If we naively take $f = f_{\rm{eff}}$, we will have $\mu \sim 10^{-1} H$. Hence, to be in the parameter region relevant to this paper, we would need $f \sim 10^{2 \sim 3} f_{\rm{eff}}$. Recently, this has been demonstrated to be achievable in a broad range of models \cite{Farina:2016tgd,Agrawal:2017cmd}.

A potential for the inflaton must be generated, making it a pseudo-Goldstone. We would like to check that a successful inflation can happen in the region of parameter space we consider in this paper. Perhaps the easiest way to satisfy such a requirement is to imagine that the decay constant $f$ dose not play a direct role in the inflaton potential. This is the case, for example, for the so-called monodromy motivated models \cite{Silverstein:2008sg,McAllister:2008hb}, which has a potential of the form
\begin{equation}
V(\phi)= W(\phi) + \Lambda^4 \cos\left( \frac{\phi}{f}\right),
\end{equation}
where $W(\phi)$, generated by stringy dynamics,  is the dominant piece in the potential which drives the inflation. For example, 
we could have  $W(\phi) = \mu^3 \phi$, with  $\phi \sim 10 \Mp$ during inflation and $\mu \sim 6 \times 10^{-4} \Mp$.  The second term  in the potential depends on $f$ and allows the possibility that the axion/inflaton couples to a confining sector. It depends on $f$. However, with the assumption $\Lambda^4 \ll W$, it does not play a significant role in driving inflation. 

We could also consider more ``economical" cases. One such example is  the so called Natural Inflation scenario \cite{Freese:1990rb}, with a potential of the form $V(\phi) = \Lambda^4 \cos (1- \cos (\phi/f_{\rm{inf}}))$, where $f_{\rm{inf}} $ is an effective decay constant and characterizes the period of the inflaton potential.  $f_{\rm{inf}}$ depends on both the symmetry breaking scale $f$ and the mechanism of generating the potential. However, for the simplest case $f_{\rm{inf}} \simeq f < \Mp$, this potential is not consistent with observations as it would produce a very red spectrum. A number of mechanisms, typically involving multiple axion-like particles, have been invented to generate an effective decay constant $f_{\rm{inf}} > \Mp > f $  \cite{Kim:2004rp,Dimopoulos:2005ac,Kaplan:2015fuy}. If we still keep $f \sim 10^{16}$ GeV as our benchmark, it is not difficult to imagine that the period of the inflaton can be larger by a factor of $10^{3\sim4}$ through one of these mechanisms, and hence this can be a viable inflation model up to potential constraint from weak-gravity arguments \cite{Rudelius:2015xta,Heidenreich:2015wga,Heidenreich:2019bjd}. We will then also invoke one of the mechanisms to enhance the coupling between the inflaton and the spectator gauge field $F$, as discussed above.

As a concrete example, we could have a  scenario in which the inflaton can have the following couplings
\begin{equation}
 \ld \supset \frac{a}{16 \pi^2} \frac{1}{M f}  \phi G\wt G + \frac{b}{16 \pi^2} \frac{N}{f} \phi F\wt F,
\end{equation}
where $a,b \sim {\mathcal{O}}(1)$. $M,N \gg 1$ are  numbers which depends on the details of the physics (at a scale $\sim f$) which generates such couplings. For example, in the models presented in Ref.~\cite{Agrawal:2017cmd}, we have $N \sim M^2$. $G$ is the field strength of a gauge field which would become strongly coupled at a scale around $\Lambda_{\rm{inf}}$, and generates a potential for the inflaton $\phi$. In this case, $H \sim \Lambda_{\rm{inf}}^2/\Mp \sim 10^{13}$ GeV. Hence, $\dot\phi_0^{1/2} \simeq 60 H \sim 10^{15}$ GeV. For a benchmark value of scale $f$, we can take $f \sim($several$) \times 10^{16} - 10^{17}$ GeV ($> \Lambda$). Therefore, $M \sim 100$ gives $f_{\rm{inf}} = M f > \Mp$. At the same time, the effective coupling to a spectator gauge field $F$ is $f_{\rm{eff}} = f/N \sim 10^{13}$ GeV.  In this case, the effective chemical potential for gauge field $F$ is $\mu \sim \dot \phi_0 / (16 \pi^2 f_{\rm{eff}}) \sim 10^{15} $ GeV. These are of course rough and  parametric estimates, and it is easy to get one order of magnitude either way.

Next, we discuss the mass of the gauge boson of the spectator gauge group. It is easy to imagine it to be Higgsed. In principle, depending on the Higgs potential, its mass can be a free parameter. For the scenario under consideration in the paper, it would be more interesting to consider the case in which the physics at the inflation scale $\Lambda_{\rm{inf}}$ is also responsible for the generation of the gauge boson mass. For example,  some strong dynamics can generate the inflation scale, such as in the models mentioned above. At the same time, the strong dynamics can also have a richer structure. It can break a global symmetry which give rise to a coset worth of Goldstones, similar to the case of QCD. With additional explicit breaking, these Goldstones can develop a potential which in the end Higgses  the spectator gauge group. Such a set up has been explored extensively for the electroweak symmetry breaking, known as the composite Higgs models (see \cite{Panico:2015jxa} for a review). Implementing a similar mechanism to our setup with the strong coupling scale being $\Lambda_{\rm{inf}}$,   we expect the gauge boson mass to be 
\begin{equation}
m_A = \frac{g}{4 \pi} \Lambda_{\rm{inf}} \sim 10^{14 \sim 15} \rm{\ GeV}. 
\end{equation}

An operator at dimension-6, 
\begin{equation}
\FR{c}{\Lambda_{\Sigma}^2}(\pd_\mu\phi)^2|\Sigma|^2,
\end{equation}
will also play an important role in our study, both giving an important contribution to the size of the Higgs VEV $\sigma_0$ and providing an important coupling to mediate the signal.  The size of $\Lambda_{\Sigma}$ is determined by the physics which mediates the interaction between the Higgs and the inflaton. Similar to the discussion of other mass scales above, it would be more natural to consider that $\Lambda_{\Sigma}$ is also related to the physics which governs the inflation. For example, we can consider the strong coupling example again, where the strong coupling scale is $\Lambda_{\rm{inf}} \sim 10^{16}$ GeV. The coupling between a pseudo-Goldstone ($\Sigma$) and the inflaton (not part of the composite states) can be mediated by one of the composite resonances with mass $m_*$ and coupling $g_*$. In general, we expect $m_* \sim g_* \Lambda_{\rm{inf}}/(4 \pi)$. Hence, we can estimate $\Lambda_{\Sigma} \sim m_* /g_* \sim \Lambda_{\rm{inf}}/(4 \pi)$.

\section{The Signal}
\label{sec_signal}

\subsection{Gauge Boson Production}

The gauge boson production from the axion coupling has been studied a lot in various contexts. Here we briefly summarize the result relevant to our investigation of cosmological collider signals. 

We begin with evaluating the Lagrangian (\ref{lag}) with the scalar background $\la\pd_\mu\phi\ra=\dot\phi_0\de_{\mu 0}$ and $\la \Sigma\ra=\si_0/\sqrt{2}$, and keeping gauge boson terms only,
\bge
  \ld\supset -\FR{1}{4}F^2-\mu t F\wt F-\FR{1}{2}a^2(\tau)m_A^2A^2,
\ede
where again $\mu=\dot\phi_0/\Lambda_F$, $m_A=g\si_0$. The physical time $t$ and the conformal time $\tau$ during inflation are related by $e^{Ht}=a=-1/(H\tau)$. Imposing the condition $\pd_\mu(\sqrt{-g}A^\mu)=0$  removes the $A_0$ component and then yields the equation of motion for the spatial component $A_i$, which can be written in terms of $k$-mode as
\begin{equation}
  A_i''+k^2 A_i+a^2m_A^2A_i-\ii a\mu\ep_{ijk}k_i A_k=0,
\end{equation}
where the prime denotes the conformal time derivative. These three equations can be decoupled by going to the helicity basis $A_{(h)}~(h=-1,0,+1)$ where $A_{(\pm 1)} = \frac{1}{\sqrt 2} (A_1\pm \ii A_2)$ and $A_{(0)}=A_3$.\footnote{More precisely, the $3$-component of the gauge field $A_3$ is not the full longitudinal polarization $A_{(0)}$. The latter also has a nonzero temporal component. But the temporal component is not independent and is related to $A_3$ by the constraint $\pd_\mu(\sqrt{-g}A^\mu)=0$, so we can discard it for now.} Then the decoupled equations read,
\begin{align}
\label{Aeom}
  A_{(h)}''+(k^2 +a^2m_A^2 + 2a\mu h k )A_{(h)}=0.
\end{align}
As explained in \cite{Wang:2019gbi}, the combination $\mu h$ is a general feature of the chemical potentials that can enhance particle productions during inflation. The dispersion relation for $A_{(\pm 1)}$ is
\begin{align}
\label{eq:dispersion}
\omega^2_{\pm}= (k_{\rm{phys}} \pm \mu)^2 + m_A^2 - \mu^2 - \frac{H^2}{4},
\end{align}
where $k_{\rm{phys}} = k/a$ is the physical momentum. Working in the limit $m_A \sim \mu \gg H$, we can ignore the last term. 
From (\ref{eq:dispersion}),  we see that the new feature for the gauge boson's chemical potential, as opposed to the fermionic case, is that the mode equation (\ref{Aeom}) can become truly tachyonic when $\mu>m_A$. As a result, both the positive and the negative frequency part of the mode function in the late-time limit will be exponentially enhanced. 

Assuming a Bunch-Davis-like initial condition for $A_{(h)}$, the solutions to equations (\ref{Aeom}) are 
\begin{align}
\label{Amode}
  A_{(h)} (\tau,\mb k)=\FR{e^{-\pi h\wt\mu/2}2^{-\ii h\wt\mu}}{\sqrt{2k}}\text{W}_{\ii h\wt\mu,\ii\wt\nu}(2\ii k\tau).
\end{align}
Here $\text{W}_{\ka,\mu}(z)$ is the Whittaker's function, $\wt\mu\equiv \mu/H$ and $\wt\nu\equiv -\ii\sqrt{1/4-(m_A/H)^2}$. We only consider $m_A/H\gg 1$ in this work so $\wt\nu$ is always real and positive. One can readily check that the early-time limit ($\tau\to-\infty$) has the following properly normalized form,
\begin{align}
  A_{(h)}(\tau,\mb k)\sim \FR{1}{\sqrt{2k}}(-k\tau)^{\ii h\wt\mu}e^{-\ii k\tau}.
\end{align}
On the other hand, the particle production can be most easily seen by looking at the late-time limit  $|k\tau|\ll 1$ of the wavefunction of the gauge boson
\begin{align}
\label{A_mode_late}
  A_{(h)}(\tau,\mb k)\sim (-\tau)^{1/2}\bigg[ e^{\ii(\wt\nu-h\wt\mu)\log 2-\ii\pi/4}\FR{e^{\pi(\wt\nu-h\wt\mu)/2}\Gamma(-2\ii\wt\nu)}{\Gamma(\fr{1}{2}-\ii h\wt\mu-\ii\wt\nu)}(-k\tau)^{\ii\wt\nu}+(\wt\nu\to-\wt\nu)\bigg].
\end{align}
On the right hand side we already see a factor $e^{-\pi h\wt\mu/2}$. It leads to exponential enhancement of mode with $h\wt\mu<0$, suppress the mode with $h\wt\mu>0$, and has no effect on the longitudinal mode $h=0$. The $h\wt\mu$ dependence in the $\Gamma$ function will be discussed below.

\begin{figure}[tbph]
\centering
\includegraphics[width=0.45\textwidth]{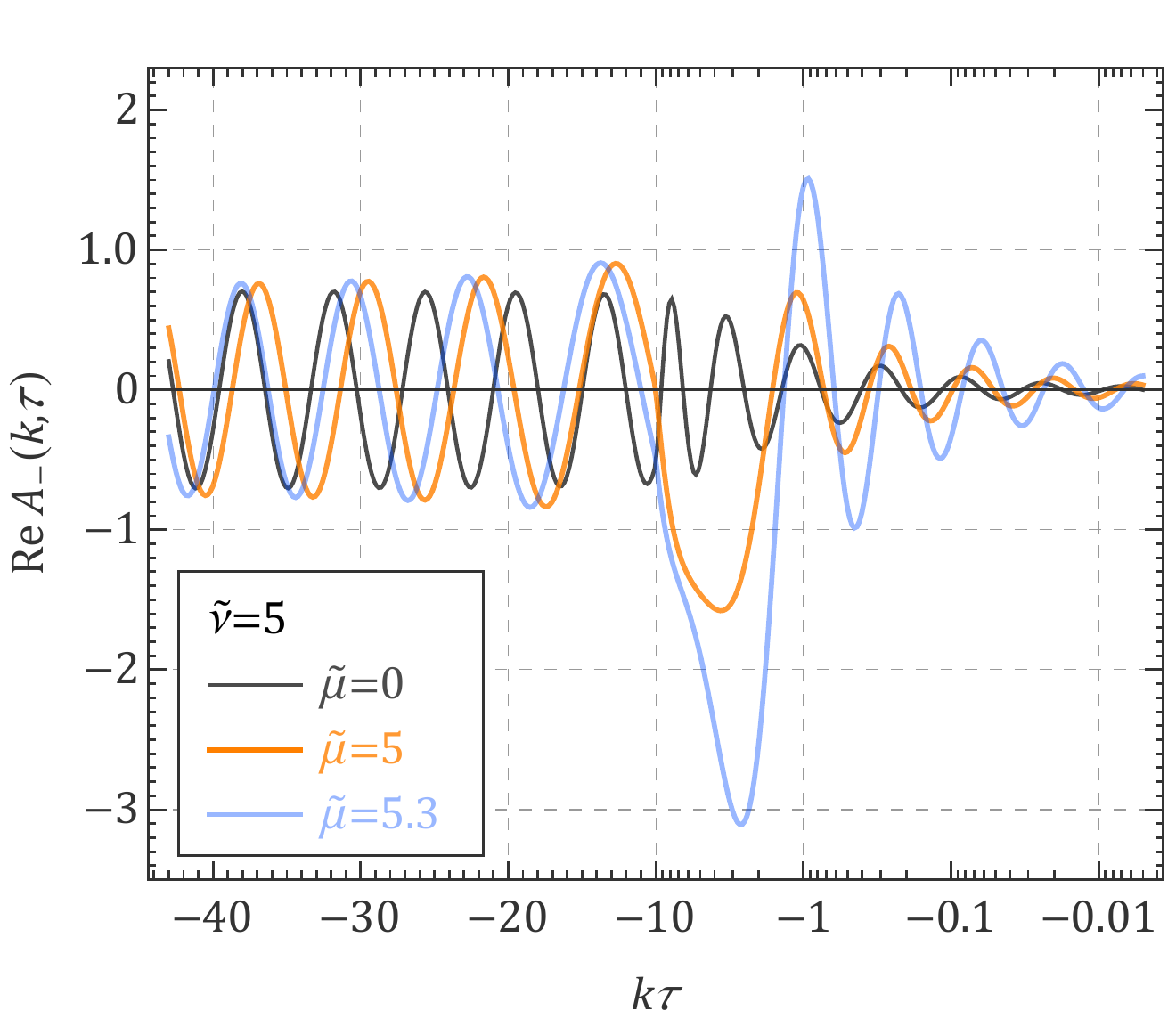}
\includegraphics[width=0.45\textwidth]{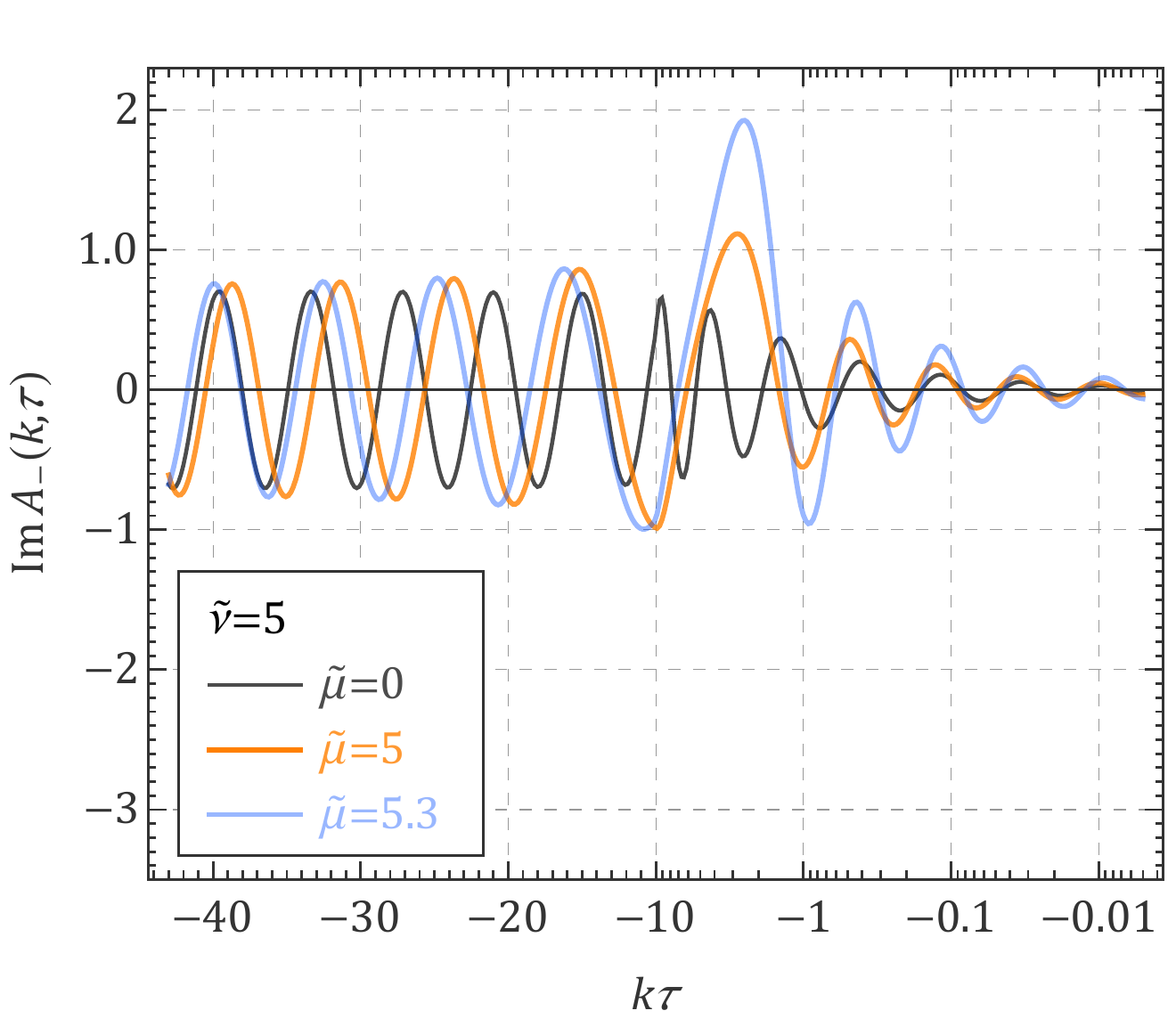}\\
\includegraphics[width=0.45\textwidth]{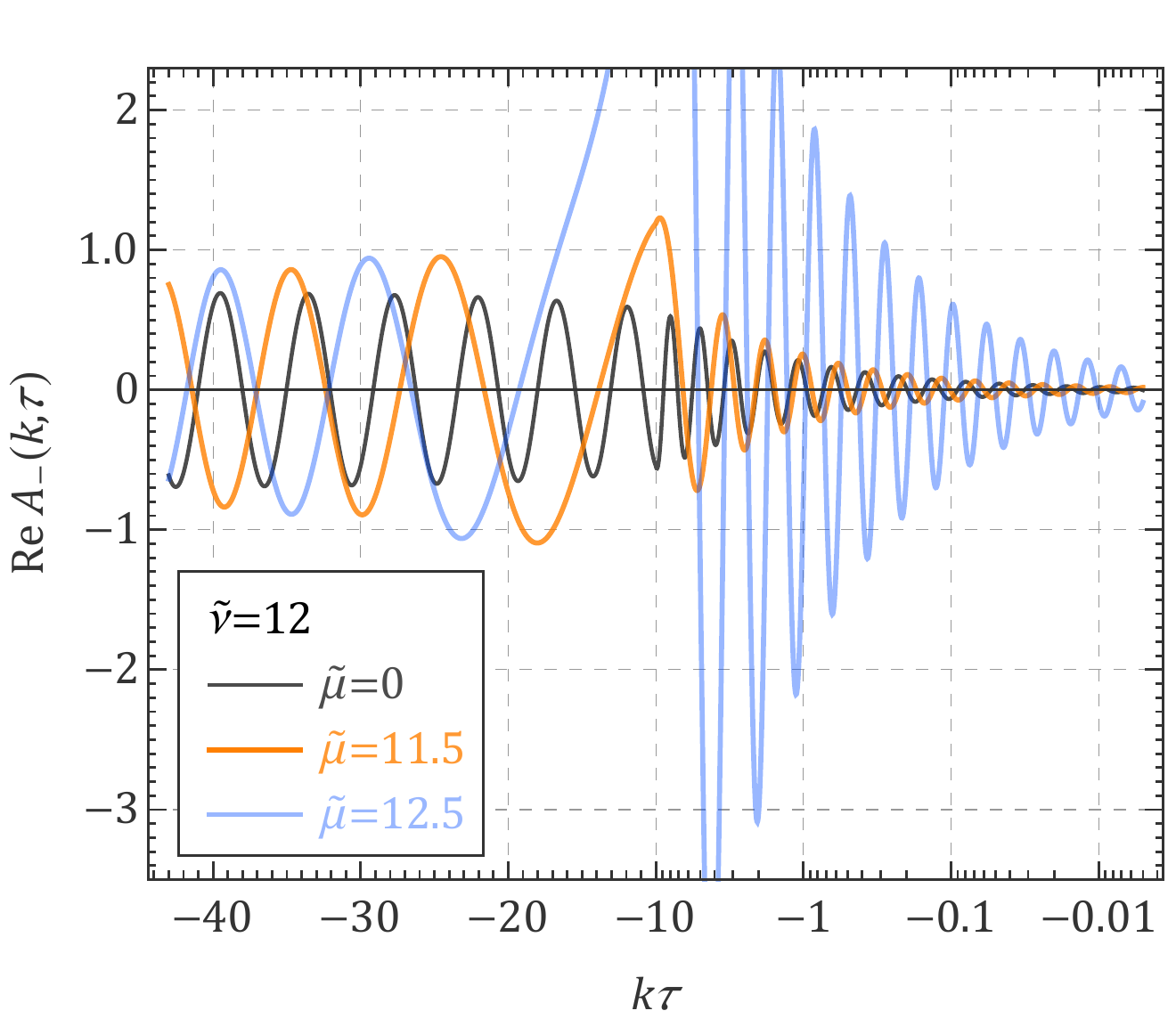}
\includegraphics[width=0.45\textwidth]{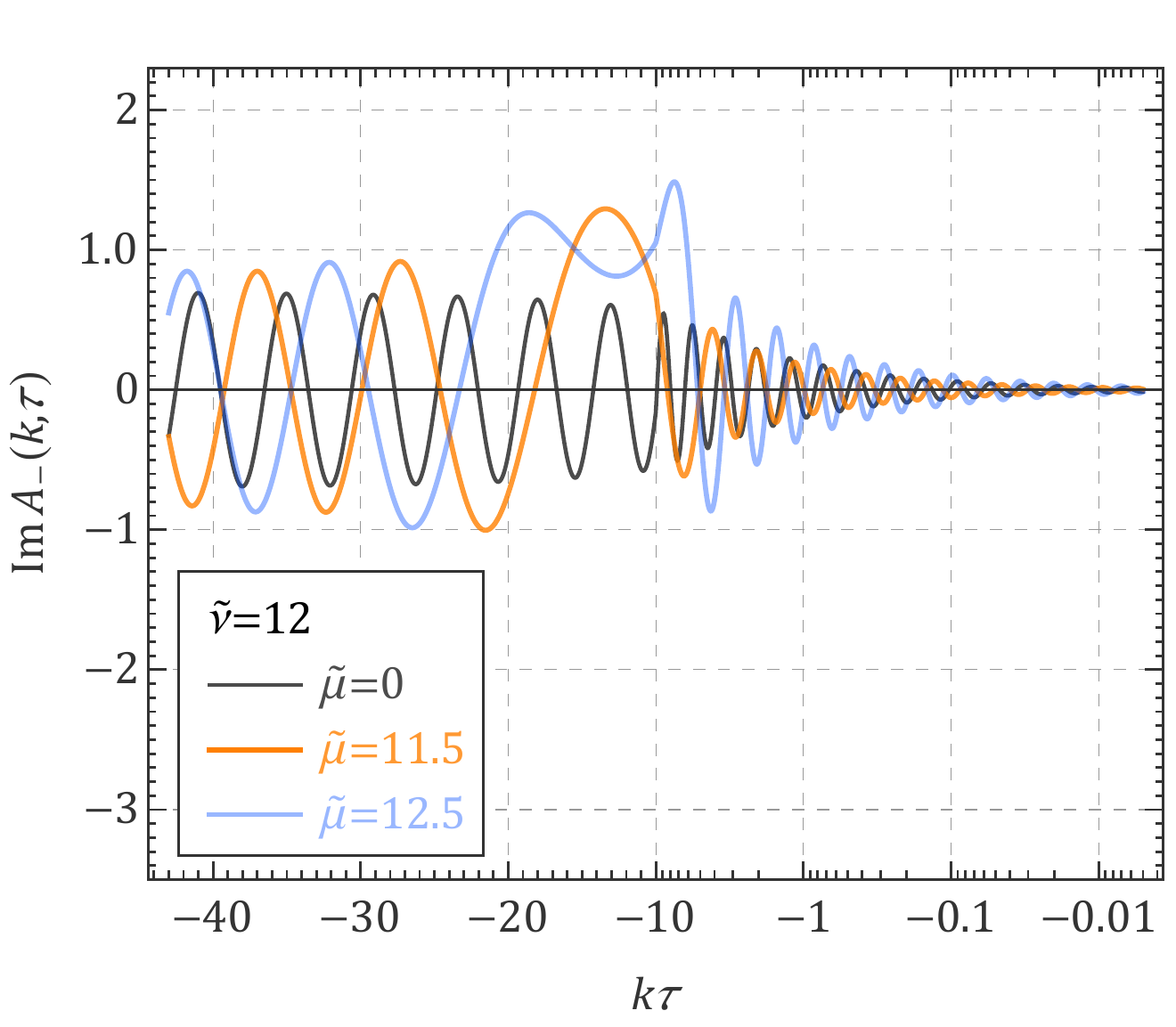}
\caption{The mode functions of the gauge boson with chemical potential in inflation. We show the real (the two left panels) and imaginary parts (the two right panels) for two different choices of mass parameter $\wt\nu=5$ (upper) and 12 (lower). In each panel, the three curves show the mode functions with three choices of chemical potential $\mu$. To properly illustrate the linear oscillation in the early time $|k\tau|\gg 1$ and logarithmical oscillation in the late time $|k\tau|\ll 1$, in each panel, we use linear scale for $k\tau<-10$ and logarithmical scale for $k\tau>-10$.  }
\label{fig_mode}
\end{figure}

Roughly speaking, the mode function receives most of its enhancement when the physical momentum $k\tau\simeq \wt\mu$ where the adiabatic approximation is maximally violated. The produced gauge bosons then  follow the comoving dilution in the late-time limit. In Fig.\;\ref{fig_mode} we show this behavior by plotting the mode function $A_{(-1)}$ in (\ref{Amode}) with fixed $k=1$ against the conformal time $k\tau$. We show the mode functions for different choices of the mass $\wt\nu$ and the chemical potential $\wt\mu$. One can see that the early-time behavior as oscillations in $k\tau$ with amplitudes and frequency independent of either $\wt\mu$ or $\wt\nu$. At late times, the mode functions develop oscillations in $\log|k\tau|$, with the frequency determined by $\wt\nu$, and the amplitude determined by both $\wt\nu$ and $\wt\mu$. (Note the change of $k\tau$ coordinate from linear scale to logarithmical scale at $k\tau=-10$.) The  enhancement of the amplitude at later times is evident for large $\wt\mu$.

\newpage
\subsection{Approximating the propagators}

Later in this paper we will calculate the oscillatory signals from these gauge boson modes. However,  it will be helpful first to estimate the signal size before a detailed calculation. For this purpose we use the fact that the signal is contributed mostly by the late-time part of the mode function, and thus will take a look at the late-time expansion of the gauge boson's propagator. 

The propagator is constructed following the standard Schwinger-Keldysh (SK) formalism \cite{Chen:2017ryl},
\bge
\label{Dgreater}
D_>^{(h)}(\mb k,\tau_1,\tau_2)=A_{(h)}(\tau_1,\mb k)A_{(h)}^*(\tau_2,\mb k).
\ede
At late times $|k\tau|\ll 1$, we can decompose the propagator into the ``local'' part and ``nonlocal'' part, where the latter contains non-integer dependence of the momentum.
\begin{align}
\label{Dgreater_nonlocal}
  &D_>^{(h)}\big|_\text{nonlocal}=(\tau_1\tau_2)^{1/2}\bigg[\FR{e^{-\pi h\wt\mu}\Gamma^2(-2\ii\wt\nu)}{\Gamma(\fr{1}{2}-\ii h\wt\mu-\ii\wt\nu)\Gamma(\fr{1}{2}+\ii h\wt\mu-\ii\wt\nu)}(4k^2\tau_1\tau_2)^{+\ii\wt\nu}+(\wt\nu\to-\wt\nu)\bigg],\\
  &D_>^{(h)}\big|_\text{local}
=\FR{(\tau_1\tau_2)^{1/2}}{2\wt\nu}\bigg[\FR{1+e^{-2\pi(\wt\nu+h\wt\mu)}}{1-e^{-4\pi\wt\nu}}\Big(\FR{\tau_1}{\tau_2}\Big)^{+\ii\wt\nu}
-\FR{1+e^{2\pi(\wt\nu-h\wt\mu)}}{1-e^{+4\pi\wt\nu}}\Big(\FR{\tau_1}{\tau_2}\Big)^{-\ii\wt\nu}\bigg].
\end{align}
Using the asymptotic expression of $\Gamma$-function $\log\Gamma(z)\sim(z-\fr{1}{2})\log z-z+\fr{1}{2}\log2\pi$, or more explicitly,
\bge
  \Gamma(a\pm \ii b)\sim \sqrt{2\pi}b^{a-1/2}e^{-\pi b/2}e^{\pm\ii(b\log b+\pi a/2)}, ~~~~a,b\in\mathbb{R}, ~~~~b\gg 1,
\ede
 we can find an expression for non-local propagator for large $\wt\mu$ and $\wt\nu$. Note that the asymptotic behavior holds well even with mildly large $|b|\gtrsim 1$, so we can apply it to, e.g., $\Gamma(\fr{1}{2}+\ii\wt\mu-\ii\wt\nu)$ where the two large numbers $\wt\mu$ and $\wt\nu$ are cancelling each other. Now we assume $\wt\mu>0$ without loss of generality, then the $h=-1$ component is dominant, with the following propagators.
\begin{eqnarray}
&&D_>^{(-)}\big|_\text{nonlocal}=\FR{(\tau_1\tau_2)^{1/2}}{2\wt\nu}\,2\text{Re}\,\Big[e^{\ii\varphi(\wt\mu,\wt\nu)}(4k^2\tau_1\tau_2)^{+\ii\wt\nu}\Big]\times\left\{\begin{split}&e^{2\pi(\wt\mu-\wt\nu)}&&(\wt\mu>\wt\nu)\\&e^{-\pi(\wt\nu-\wt\mu)} &&(\wt\mu<\wt\nu)\end{split}\right.\\
&&D_>^{(-)}\big|_\text{local}=\FR{(\tau_1\tau_2)^{1/2}}{2\wt\nu}\times\left\{\begin{split}&-2e^{2\pi(\wt\mu-\wt\nu)}\cos\Big(\wt\nu\log\FR{\tau_1}{\tau_2}\Big)&&(\wt\mu>\wt\nu)\\&\Big(\FR{\tau_1}{\tau_2}\Big)^{+\ii\nu}+e^{-2\pi(\wt\nu-\wt\mu)}\Big(\FR{\tau_1}{\tau_2}\Big)^{-\ii\wt\nu} &&(\wt\mu<\wt\nu)\end{split}\right.
\end{eqnarray}
From these results we can read a rule for simple estimate. To summarize this rule more compactly, we introduce the following two quantities,
\bge
  \ga\equiv \wt\mu-\wt\nu\simeq (\varrho-1)m_A/H;~~~~~~
  \Omega(\ga)\equiv\left\{
  \begin{split}
  &\ga,~~~~&&(\ga\geq0)\\
  &0. &&(\ga<0)
  \end{split}
  \right.
\ede
Then the estimate goes as
\bge
\label{Dest}
\boxed{~~~~\phantom{\bigg(}
\begin{aligned}
  &D_>\big|_\text{nonlocal}\simeq  e^{\pi[\ga+\Omega(\ga)]},
  &&D_>\big|_\text{local}\simeq \wt\nu^{-1}e^{2\pi \Omega(\ga)}.
\end{aligned}
~~~~\phantom{\bigg)}}
\ede
Here we have dropped a prefactor $\wt\nu^{-1}$ in estimating the nonlocal part of $D_>$, since this factor is usually compensated by a positive power $\wt\nu$ coming from the in-in integral. In the large mass limit, $\wt\nu \sim m_A/H$, $D_>\big|_\text{local} \propto m_A^{-1} e^{2\pi \Omega(\ga)} $. As discussed in App.\ \ref{app_mass_dep}, the time integral will change the mass dependence  to $m_A^{-2}$ and thus reproduce the EFT limit. At the same time, the exponential factor $e^{2\pi \Omega(\ga)}$ is still present in this limit as the particle production only depends on the relative size of $\mu$ and $m_A$.  Hence, in the large $m_A$ limit, we can approximate a hard propagator as $m_A^{-2} e^{2\pi \Omega(\ga)}$.

\subsection{Estimate of Signal Size}
\label{sec_est}

\begin{figure}[tbph]
\centering
\includegraphics[width=0.98\textwidth]{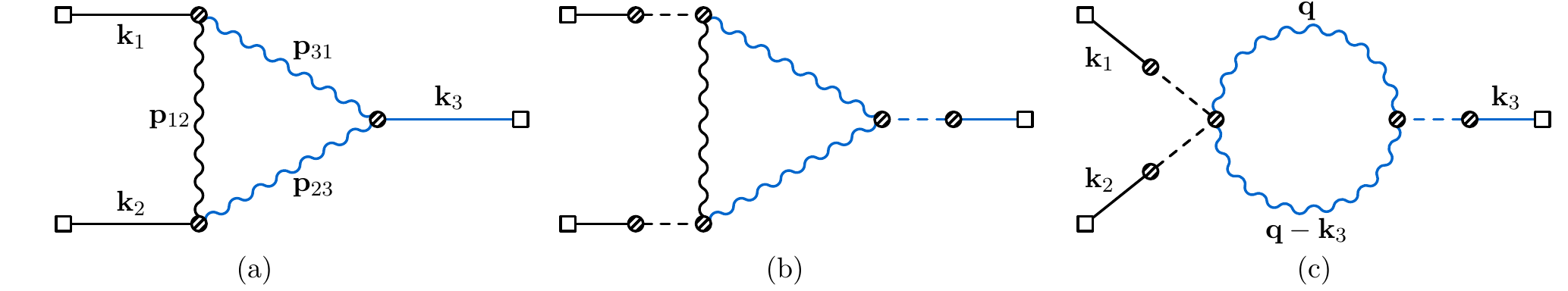}
\caption{One-loop diagrams contributing to gauge boson signals. The blue color marks the soft lines in the squeezed limit.}
\label{fig_loop}
\end{figure}

With the rule of approximating gauge boson propagator as derived above, we can now estimate the size of gauge boson signals in NG. Different from the fermion case where there is only one relevant 1-loop diagram, here we have at least two relevant categories of diagrams to consider, shown in Fig.\;\ref{fig_loop}.

In the first category shown in Fig.\;\ref{fig_loop}(a), the gauge boson loop is directly attached to the inflaton external lines, with the coupling from $\phi F\wt F$. In the second category shown in Fig.\;\ref{fig_loop}(b) and (c), we mix the Higgs $\si$ (dashed lines) and the inflaton (solid lines) via the dim-6 operator. The mixing is of order $\dot\phi_0\si_0/\Lambda_\Sigma^2$. Both $\dot\phi$ and $\si_0$ could be large so this mixing could be large, too. There are suppressions from Higgs internal lines $\sim 1/m_\si^2$ but this can well be compensated by the two-point mixing as $u \to 1$. Of course there can be mixed case where the gauge boson loop is attached to both $\si$ and $\de\phi$. For now we will only focus on the diagrams in Fig.\;\ref{fig_loop}. 

We will now estimate the NG of each diagram. We will consider both the non-signal part of the NG, which we refer to as the ``background,'' and the oscillatory signals. The background estimation is useful when translating the current NG constraints to that of model parameters. The way of estimating NG can be roughly summarized as ``loop factors $\times$ propagators $\times$ vertices'' multiplied by a numerical factor $1/(2\pi P_\zeta^{1/2})$ where $P_\zeta\simeq 2\times 10^{-9}$ is the amplitude of the scalar power spectrum. We refer readers to \cite{Wang:2019gbi} for a more detailed description about the signal estimate. We note that there are known subtleties about this estimate. In particular, we can use this rule to figure out easily the exponential dependence on the model parameters (the chemical potential and the mass), but generally we cannot get the correct power dependence. In fact, we don't even know how to calculate this power dependence analytically in general situation. We discuss these subtleties in more detail in App.\ \ref{app_mass_dep}.

Now we estimate the signal size for each diagram. To simplify our expressions, we will take the unit $H=1$ in the rest of this subsection. We will also focus on large mass region $\wt\nu\gg 1$ so that $\wt\nu\simeq m_A$. 
Our final results depend only on three free parameters $\varrho\equiv \mu/m_A$, $m_A$, and $u$ introduced in (\ref{par}). The diagram in Fig\;\ref{fig_loop}(a) can be estimated as
\begin{align}
\label{f_a_bg_est}
  f_\text{NL}^{\text{(bg)}}(\text{a})\sim &~ 
  \FR{1}{16\pi^2}\FR{1}{2\pi P_\zeta^{1/2}}\FR{1}{\Lambda_F^3}\FR{1}{m_A^6}e^{6\pi\Omega(\ga)}=\FR{1}{4}P_\zeta\varrho^3m_A^{-3}e^{6\pi\Omega(\ga)},\\
  f_\text{NL}^{\text{(osc)}}(\text{a})
  \sim&~  \FR{1}{16\pi^2}\FR{1}{2\pi P_\zeta^{1/2}}\FR{1}{\Lambda_F^3}\FR{1}{m_A^2}e^{2\pi[\ga+2\Omega(\ga)]}
  =\FR{1}{4}P_\zeta\varrho^3m_A e^{2\pi[\ga+2\Omega(\ga)]}.
\end{align}
We have included a loop factor $1/16\pi^2$, the prefactor $1/(2\pi P_\zeta^{1/2})$, three vertices, each of which gives $1/\Lambda_F$, and finally the internal gauge boson propagators  taken from (\ref{Dest}). We estimate two types of NG.  $ f_\text{NL}^{\text{(bg)}}$ is obtained by approximating each of the internal gauge boson propagators with the local form $m_A^{-2} e^{2\pi \Omega(\ga)}$. This is the kind of NG which will be constrained by the current observation.  $f_\text{NL}^\text{(osc)}$ is the size of the oscillatory signal. To estimate this, we approximate the hard line (black in Fig.\;\ref{fig_loop}) with the local form propagator, while the soft lines (blue in Fig.\;\ref{fig_loop}) should take the non-local propagator in (\ref{Dest}).  

The sizes of the rest of the diagrams in Fig.\;\ref{fig_loop} can be estimated in a similar way. For Fig.\;\ref{fig_loop}(b):
\begin{align}
\label{f_b_bg_est}
  f_\text{NL}^{\text{(bg)}}(\text{b})\sim &~ \FR{1}{16\pi^2}\FR{1}{2\pi P_\zeta^{1/2}}\bigg(\FR{\dot\phi_0\si_0}{\Lambda_\Sigma^2}\FR{1}{m_\si^2}\FR{1}{m_A^2}(2g^2\si_0)\bigg)^3e^{6\pi\Omega(\ga)}=\FR{1}{4}P_\zeta u^{-3} e^{6\pi\Omega(\ga)},\\ 
\label{f_b_osc_est}
  f_\text{NL}^{\text{(osc)}}(\text{b})\sim &~ \FR{1}{16\pi^2}\FR{1}{2\pi P_\zeta^{1/2}}\bigg(\FR{\dot\phi_0\si_0}{\Lambda_\Sigma^2}\FR{1}{m_\si^2}(2g^2\si_0)\bigg)^3\FR{1}{m_A^2}e^{2\pi[\ga+2\Omega(\ga)]}=\FR{1}{4}P_\zeta u^{-3}m_A^4 e^{2\pi[\ga+2\Omega(\ga)]}.
\end{align}
Here we have taken the $\si$-propagators to be $1/m_\si^2$ since we always focus on $m_\si\gg 1$. Finally, Fig.\;\ref{fig_loop}(c) is
\begin{align}
\label{f_c_bg_est}
  f_\text{NL}^{\text{(bg)}}(\text{c})\sim &~\FR{1}{2}\FR{1}{16\pi^2}\FR{1}{2\pi P_\zeta^{1/2}}\bigg(\FR{\dot\phi_0\si_0}{\Lambda_\Sigma^2}\FR{1}{m_\si^2} \bigg)^3(2g^2)(2g^2\si_0)\FR{1}{m_A^4}e^{4\pi\Omega(\ga)}=\FR{1}{16}P_\zeta u^{-3} e^{4\pi\Omega(\ga)} ,\\ 
\label{f_c_osc_est}  
  f_\text{NL}^{\text{(osc)}}(\text{c})\sim &~\FR{1}{2} \FR{1}{16\pi^2}\FR{1}{2\pi P_\zeta^{1/2}}\bigg(\FR{\dot\phi_0\si_0}{\Lambda_\Sigma^2}\FR{1}{m_\si^2}\bigg)^3(2g^2)(2g^2\si_0)e^{2\pi[\ga+\Omega(\ga)]}=\FR{1}{16}P_\zeta u^{-3} m_A^4 e^{2\pi[\ga+\Omega(\ga)]} .
\end{align}

To get an idea of overall signal strength and also the relative importance of difference diagrams, we can look at a special case where $\varrho\simeq 1$ and $m_A\gg1$. This is the parameter range we are mostly interested in. One might want to consider the case with $\varrho>1$ where the large exponential factor can lead to great enhancement. However, a large exponential factor could be severely constrained by several physical considerations as we will elaborate in the next section. Therefore we will take those exponential factors as $\order{1}$ for now. Then, we see that each diagram is simply a factor of $P_\zeta$ multiplied by some powers of $m_A$, with an expected range $1\ll m_A\lesssim \dot\phi_0^{1/2}\simeq 60$, as well as by a factor of $u^{-3}$ for Diagrams (b) and (c). The factor $P_\zeta\simeq 10^{-9}$ would make the signal tiny, unless there is a large positive power of $m_A$ or if $u\ll 1$. We see that Diagram (a) is independent of $u$, and the signal $f_\text{NL}^\text{(osc)}$ has only one positive power in $m_A$. Therefore Diagram (a) will be tiny in any case. On the other hand, (b) and (c) get more powers of $m_A$. So they will be the dominant contribution when $u\sim 1$ or $u<1$, although they will also be suppressed when $u\gg 1$. We note that this result is independent of the relative size between $\Lambda_F$ and $\Lambda_\Sigma$ as long as all parameters are within the range of validity of our estimate. (In particular, we must have $m_A>H$ for our estimate to be valid. Therefore we require $\si_0/H>1/g$.)

To summarize, we find that the largest signal is from Diagrams (b) and (c). Even without a truly exponential enhancement, we are able to get large signals with the help of the factor $m_A^4$, although we should note again that it is generally difficult to estimate to power dependence correctly. Indeed, a more careful calculation in the following section shows that this power dependence is actually $m_A^{11/2}$ rather than $m_A^4$. But even this result may not capture the full power dependence on the mass $m_A$. We comment on this issue in the App.\;\ref{app_mass_dep} and leave a possible improved calculation for future study.

And it can be observed at this point that the large signal should survive all the constraints mentioned above. This is because all those constraints are actually constraining a large exponential factor, namely that $e^{2\pi\ga}$ cannot be much greater than 1. But here we see that we do not need a large exponential factor.

\subsection{Explicit Calculation} 
\label{sec:explicit}

The estimate above shows that Diagrams (b) and (c) can possibly give rise to visibly large signals while Diagram (a) is always tiny. Therefore we will calculate (b) and (c) more explicitly in this subsection. We will present some detailes of calculation of Diagram (b). Diagram (c) can be calculated quite similarly, to which we can simply adapt the result of Diagram (b) with appropriate changes. We follow the method in \cite{Chen:2018xck} but with improvements. More details about the formalism and the convention we used here are reviewed in \cite{Chen:2017ryl}. Uninterested readers can directly go to the final results in (\ref{fnl_b_simp}) and (\ref{fnl_c_simp}).
 
According to the diagrammatic rule, Diagram (b) can be written as
\begin{align}
\label{diag_b}
  \la\de\phi^3\ra_\text{(b)}=&\sum_{\mathsf{a}_i=\pm} \mathsf{a}_1\mathsf{a}_2\mathsf{a}_3(\ii g^2\si_0)^3\int\prod_{i=1}^3\bigg[\FR{\di\tau_i}{|H\tau_i|^2}\mathcal{G}_{\mathsf{a}_i}(k_i;\tau_i)\bigg]\n\\&\times\int\FR{\di^3\mb q}{(2\pi)^3}D_{\mathsf{a}_1\mathsf{a}_2\mu}{}^{\nu}(\mb p_{12};\tau_1,\tau_2)D_{\mathsf{a}_2\mathsf{a}_3\nu}{}^{\lam}(\mb p_{23};\tau_2,\tau_3)D_{\mathsf{a}_3\mathsf{a}_1\lam}{}^{\mu}(\mb p_{31};\tau_3,\tau_1).
\end{align}
Here $\mathsf{a}_i=\pm1$ $(i=1,2,3)$ are in-in contour indices, and the momenta are as labeled in Fig\;\ref{fig_loop}(a). $D_{\mathsf{a}_i\mathsf{a}_j\mu}{}^\nu$'s are gauge boson propagators, and we have written the external lines compactly in terms of mixed propagators \cite{Chen:2017ryl},
\bge
  \mathcal{G}_\pm(k,\tau)=\FR{2\ii\dot\phi_0\si_0}{\Lambda_\Sigma^2}\sum_{\mathsf{a}=\pm}\mathsf{a}\int\FR{\di\tau'}{|H\tau'|^{3}}\big[\pd_{\tau'}G_{\mathsf{a}}(k,\tau')\big]D_{\mathsf{a}\pm}(k;\tau',\tau),
\ede
in which $G_\mathsf{a}$ is the bulk-to-boundary propagator of the inflaton fluctuation and $D_{\mathsf{a}\pm}$ is the propagator for scalar $\sigma$. The integrals in (\ref{diag_b}) are difficult to be carried out directly, and we adopt several approximation to make progress. 

\paragraph{Approximation of the mixed propagator.} First, since the oscillation signals associated with the mixed propagator  are small, we will ignore them. Hence, we can expand the mixed propagator in the large mass limit. This is equivalent to replacing the $\si$-lines by effective vertices $1/m_\si^2$. More explicitly, we derive the effective vertex as follows.
\begin{align}
\label{2ptmixLag}
  &\int\di^4x\sqrt{-g(x)}\int\di^4y\sqrt{-g(y)}\FR{2}{\Lambda_\Sigma^2}a^{-1}(x)\dot\phi_0\si_0\de\phi'(x)\cdot g^2\si_0A_\mu(y)A^\mu(y)\la\si(x)\si(y)\ra\n\\
  =&\int\di^4x\sqrt{-g(x)}\int\di^4y\sqrt{-g(y)}\FR{2}{\Lambda_\Sigma^2}a^{-1}(x)\dot\phi_0\si_0\de\phi'(x)\cdot g^2\si_0A_\mu(y)A^\mu(y)\FR{1}{m_\si^2}\de^{(4)}(x-y)\n\\
  =&\int\di^4x\FR{1}{|H\tau|}\FR{m_A^2}{u\dot\phi_0}\de\phi'\eta^{\mu\nu}A_\mu A_\nu.
\end{align}
In the last line we used (\ref{par}), and in the second line we made the substitution $\la\si(x)\si(y)\ra\to m_\si^{-2}\de^{(4)}(x-y)$.\footnote{This is most easily justified with Euclidean dS representation \cite{Chen:2016hrz}, where the propagators can be decomposed into spherical harmonics $Y_{\vec L}(x)$,
\[ 
  \la\si(x)\si(y)\ra= \sum_{\vec L}\FR{H^2}{L(L+3)+(m_\si/H)^2}Y_{\vec L}(x)Y_{\vec L}^*(y) 
  \to~\FR{H^4}{m_\si^2}\sum_{\vec L}Y_{\vec L}(x)Y_{\vec L}^*(y)=\FR{1}{m_\si^2}\de^{(4)}(x-y).
\]
We will also present a less rigorous derivation within the current formalism in App.\ \ref{app_mass_dep}.
}
We  expect this coupling to vanish as take $\Lambda_\Sigma \to \infty$. This is indeed the case as $m_A^2 / u = g^2 \ob \sigma_0^2  \propto 1/\Lambda_\Sigma^2 $. 
In practice, however, this limit needs to be taken with care. As $\Lambda_\Sigma \gg m_\Sigma$, $m_\si^2 \simeq 2\dot\phi_0^2/\Lambda_\Sigma^2$. The EFT approximation used here breaks down when $m_\si^2\lesssim H^2$ or $\Lambda_\Sigma \gtrsim\sqrt{u}\dot\phi_0/H$.  Hence, for this calculation we assume $\dot\phi_0^{1/2}<\Lambda_\Sigma<\sqrt{u}\dot\phi_0/H$. In the unit of $H=1$, this means $60\lesssim \Lambda_\Sigma \lesssim 3600\sqrt{u}$.

\paragraph{Soft gauge boson propagator.} Next, to carry out the loop integral, we will make a late-time ($|k\tau|\ll 1$) expansion to the soft gauge boson propagators (blue lines in Fig.\;\ref{fig_loop}). For this purpose, we decompose the gauge boson propagator into components with fixed helicity, 
\begin{align}
\label{Dab}
  D_{\mathsf{ab};\mu\nu}(\mb k,\tau_1,\tau_2)=\sum_h e_\mu^{(h)}(\mb k)e_\nu^{(h)*}(\mb k)D_\mathsf{ab}^{(h)}(k,\tau_1,\tau_2),
\end{align}
and use the nonlocal part of the propagator $D_>^{(h)}$ shown in (\ref{Dgreater_nonlocal}). (The relation between $D_\mathsf{ab}^{(h)}$ and $D_{>}^{(h)}$ is reviewed in \cite{Chen:2017ryl}). 
 
 This is the crudest approximation among all we make in this calculation. There are two unsatisfactory points about it. First, the late-time expansion holds well only when $|k\tau|\lesssim1$. On the other hand the time integral in (\ref{diag_b}) receive contributions for all $|k \tau |\lesssim |\wt \mu|$ which is outside the range of validity of the expansion. Second, the local part of the propagator can actually contribute to oscillatory signals, a point that was overlooked in previous studies. We provide a detailed discussion of these issues in App.\ \ref{app_mass_dep}.

\paragraph{Hard gauge boson propagator.} For the hard loop line (the black line in Fig.\;\ref{fig_loop}), \cite{Chen:2018xck} evaluates it at the saddle point of the integral while \cite{Hook:2019zxa} used an ``improved'' late-time expansion. None of these worked perfectly. So in this work we will work in the large-mass expansion, assuming that this hard line can be approximated by an EFT vertex $1/m_A^2$. The advantage of this approximation is that it allows us to carry out the loop momentum integral completely, as opposed to previous works where the loop integral is evaluated at some particular momentum configurations. Therefore, we will make the substitution $\la A_\mu(x)A_\nu(y)\ra \to g_{\mu\nu}e^{2\pi\Omega(\ga)}m_A^{-2}\de^{(4)}(x-y)$. We have inserted an additional factor $e^{2\pi\Omega(\ga)}$ following (\ref{Dest}). This takes account of the fact that the gauge boson mode function becomes tachyonic when $\ga>0$ and the exponential enhancement applies to both the positive- and the negative-frequency parts of the mode function. Clearly we should not trust this substitution when $\ga\gg 1$.
From this we get the effective vertex,
\bge
  \int\di^4x \FR{m_A^2e^{2\pi\Omega(\ga)}}{u^2\dot\phi_0^2}(\de\phi')^2\eta^{\mu\nu}A_\mu A_\nu.
\ede
Note that we have separated all scale factors so the metric appeared here is the Minkowski metric $\eta_{\mu\nu}$ rather than the FRW metric. 

\paragraph{Simplified Result with Approximations.} Above we have listed all approximations we made in order to carry out the integral analytically. As a result, the integral (\ref{diag_b}) can be recast into the following form
\begin{align}
\label{diag_b_simp}
  \la\de\phi^3\ra_\text{(b)}\simeq &~\FR{1}{2} \FR{4\ii m_A^2e^{2\pi\Omega(\ga)}}{u^2\dot\phi_0^2}\FR{2\ii m_A^2}{u\dot\phi_0}\n\\
  &\times\sum_{\mathsf{a},\mathsf{b}=\pm}\mathsf{ab} \int \di\tau_1 \FR{\di\tau_3}{|H\tau_3|}\pd_{\tau_1}G_\mathsf{a}(k_1,\tau_1)\pd_{\tau_1}G_\mathsf{a}(k_2,\tau_1)\pd_{\tau_3}G_\mathsf{b}(k_3,\tau_3)\mathcal{I}(k_3,\tau_1,\tau_3),
\end{align}
where the loop integral $\mathcal{I}(k_3,\tau_1,\tau_3)$ is defined as
\begin{align}
\mathcal{I}(k_3,\tau_1,\tau_3)\equiv\int\FR{\di^3\mb q}{(2\pi)^3}D_{\mu\nu}(\mb q;\tau_1,\tau_3)D^{\mu\nu}(\mb q-\mb k_3;\tau_3,\tau_1).
\end{align}
Note that we have dropped the SK indices for the soft gauge boson lines. Now this has the same form as Fig.\;\ref{fig_loop}(c) after contracting the hard internal line. Working out the vertex coefficients shows
\bge
  \la\de\phi^3\ra_\text{(b)}\simeq 2e^{2\pi\Omega(\ga)}\la\de\phi^3\ra_\text{(c)}.
\ede

\paragraph{Loop integral.} Now we perform the loop integral. Without loss of generality we assume $\mu>0$ in which case the negative helicity component dominates the result. The loop integral can then be written as
\begin{align}
 \mathcal{I}(k_3,\tau_1,\tau_3)=\int\FR{\di^3\mb q}{(2\pi)^3} e_\mu^{-}(\mb q)e_\nu^{-*}(\mb q)e^{-*\mu}(\mb p)e^{-\nu}(\mb p)D^-(q,\tau_1,\tau_3)D^-(p,\tau_1,\tau_3),
\end{align}
where $\mb p\equiv\mb q-\mb k_3$.
Using rotation symmetry we can put 
\begin{align}
&\mb k_3=(0,0,k_3),
&&\mb q=(0, q\sin\theta,q\cos\theta),
&&\mb p\equiv (0,p\sin\chi,p\cos\chi)=(0,q\sin\theta,q\cos\theta-k_3).
\end{align} 
From this we find 
\begin{align}
&\mb e^-(\mb q)=\FR{1}{\sqrt 2}(1,-\ii\cos\theta,+\ii\sin\theta),
&&\mb e^{-}(\mb p)=\FR{1}{\sqrt 2}(1,-\ii\cos\chi,+\ii\sin\chi).
\end{align} 
So the polarization product in the loop integrand is $[1+\cos(\theta-\chi)]^2/4$. Then the integral can be carried out analytically. We keep the non-local part only, and the result is,
\begin{align}
  &\mathcal{I}(k_3,\tau_1,\tau_3)=\mathcal{C}(\wt\mu,\wt\nu)k_3^3\tau_1\tau_3(4k_3^2\tau_1\tau_3)^{2\ii\wt\nu}+\text{c.c.},\\
  &\mathcal{C}(\wt\mu,\wt\nu)\equiv \FR{e^{2\pi\wt\mu-4\pi\wt\nu}(4\wt\nu^4-8\ii\wt\nu^3-6\wt\nu^2+2\ii\wt\nu+3/8)\Gamma^2(-2\ii\wt\nu)}{(\ii\wt\nu)^2\Gamma(5+4\ii\wt\nu)\Gamma^2(\fr{1}{2}+\ii\wt\mu-\ii\wt\nu)\Gamma^2(\fr{1}{2}-\ii\wt\mu-\ii\wt\nu)}.
\end{align}

\paragraph{Time integral.} Now it is straightforward to finish the time integral in (\ref{diag_b_simp}), and the result is
\begin{align}
  \la\de\phi^3\ra_\text{(b)}\simeq\FR{H^5m_A^4e^{2\pi\Omega(\ga)}}{128u^3\dot\phi_0^3}\FR{1}{k_1^4k_3^2}\bigg[e^{2\pi\wt\nu}\Gamma(4+2\ii\wt\nu)\Gamma(2+2\ii\wt\nu)\mathcal{C}(\wt\mu,\wt\nu)\bigg(\FR{2k_3}{k_1}\bigg)^{2+2\ii\wt\nu}+\text{c.c.}\bigg].
\end{align}
It is conventional to represent the result in terms of the dimensionless shape function $\mathcal{S}(k_1,k_2,k_3)$, related to the correlation function $\la\de\phi^3\ra$ through
\begin{align}
  \mathcal{S}(k_1,k_2,k_3)=&-\FR{(k_1k_2k_3)^2}{(2\pi)^4P_\zeta^2}\Big(\FR{H}{\dot\phi_0}\Big)^3\la\de\phi_{k_1}\de\phi_{k_2}\de\phi_{k_3}\ra'.
\end{align}
The prime on the correlator means the $\de$-function of momentum conservation stripped, which was implicitly assumed in all previous expressions.
Therefore we have, in the squeezed limit $k_1\simeq k_2\gg k_3$,
\begin{align}
  \mathcal{S}_\text{(b)}(k_1,k_3)\simeq-\FR{\pi^2}{32}P_\zeta\FR{m_A^4}{H^4}\FR{e^{2\pi\Omega(\ga)}}{u^3}\,2\text{Re}\,\bigg[e^{2\pi\wt\nu}\Gamma(4+2\ii\wt\nu)\Gamma(2+2\ii\wt\nu)\mathcal{C}(\wt\mu,\wt\nu)\bigg(\FR{2k_3}{k_1}\bigg)^{2+2\ii\wt\nu}\bigg].
\end{align} 
Similarly,
\begin{align}
  \mathcal{S}_\text{(c)}(k_1,k_3)\simeq-\FR{\pi^2}{64}P_\zeta\FR{m_A^4}{H^4}\FR{1}{u^3}\,2\text{Re}\,\bigg[e^{2\pi\wt\nu}\Gamma(4+2\ii\wt\nu)\Gamma(2+2\ii\wt\nu)\mathcal{C}(\wt\mu,\wt\nu)\bigg(\FR{2k_3}{k_1}\bigg)^{2+2\ii\wt\nu}\bigg].
\end{align} 
It is useful to show the limit of this shape function when $\mu\sim m_A\gg H$. Using $\varrho=\mu/m_A\simeq 1$ and $\wt\nu\simeq m_A/H$, we find,
\begin{align}
\label{fnl_b_simp}
  f_\text{NL}^\text{(osc)}(\text{b})\simeq &~\FR{\pi^{3/2}}{128\sqrt{2}}P_\zeta  \Big(\FR{m_A}{H}\Big)^{11/2}\FR{e^{2\pi[\ga+2\Omega(\ga)]}}{u^3},\\
\label{fnl_c_simp}
  f_\text{NL}^\text{(osc)}(\text{c})\simeq &~\FR{\pi^{3/2}}{256\sqrt{2}}P_\zeta  \Big(\FR{m_A}{H}\Big)^{11/2}\FR{e^{2\pi[\ga+\Omega(\ga)]}}{u^3}.
\end{align}
This agrees with the previous estimates (\ref{f_b_osc_est}) and (\ref{f_c_osc_est}) in overall parametric dependences, except for the power dependence in $m_A$. The calculation here yields $m_A^{11/2}$ while the previous estimate gives $m_A^4$. From the above calculation it is clear that the additional powers $m_A^{3/2}$ are generated during performing the loop integral and the time integral, which can in no way be easily estimated.

\section{Constraints}
\label{sec_constraints}

Unlike the chemical potential for fermion production studied previously, the gauge boson signal obtained here in (\ref{fnl_b_simp}) and (\ref{fnl_c_simp}) can be exponentially large if naively extrapolated to large $\varrho\gg 1$. Clearly we should not trust the result with a large exponential enhancement as the perturbation theory breaks down there. In addition, with the presence of exponential enhancement, we will also meet several physical constraints. In this section we will outline the constraints in the parameter space. 

First of all, there is a constraint from the validity of EFT expansion used in the Lagrangian in Eq.(\ref{lag}). To make sure the derivative expansion in $\pd\phi$ is valid when evaluated with the inflation background, we require
\bge
\label{eft_con}
\Lambda_{\rm{\Sigma}}, \ \Lambda_F >\dot\phi_0^{1/2}\simeq 60H.
\ede

In addition to the above EFT constraint, there are three constraints we will consider. First, the energy density of the produced gauge bosons should be subdominant during the inflation. Second, the equilateral non-Gaussianity should be within the current limit. Third, the perturbative expansion should be justified for our computation of signals to be valid. Basically all these constraints require that the chemical potential not to be greater than the mass of the gauge boson. But there can be slight difference in numerical factors in each case which we shall go over below.

\paragraph{Energy density.} One physical constraint is that the gauge boson must not dominate the energy density, $\rh_A\ll 3\Mp^2H^2$. That is, we are assuming a conventional inflation scenario where the exponential expansion is driven by the inflaton's potential energy. This is not necessary. In fact one can consider a scenario ``warm inflation'' from such gauge boson production \cite{Anber:2009ua,Anber:2012du,Berghaus:2019whh}. We will leave this for a future study. 

The energy density $\rh_A$ of produced gauge boson can be found from the mode function in the late-time limit (\ref{A_mode_late}). 
This late-time behavior can be compared with the normalized basis
\bge
  u(\tau,\mb k)\sim \FR{1}{\sqrt{2\wt\nu}}\al_h e^{\ii\varphi_1}(-\tau)^{1/2+\ii\nu}+\FR{1}{\sqrt{2\wt\nu}}\be_h e^{\ii\varphi_2}(-\tau)^{1/2-\ii\nu}
\ede
From this we can read the Bogoliubov coefficient $\be_h$ as
\bge
  \be_h=\FR{e^{-\pi(\wt\nu+h\wt\mu)/2}\sqrt{2\wt\nu}\Gamma(2\ii\wt\nu)}{\Gamma(\fr{1}{2}-\ii s\wt\mu+\ii\wt\nu)}.
\ede
Then, the number density $n_k$ of produced gauge boson in the phase space is given by
\bge
  n_k=|\be_h|^2=\FR{e^{2\pi(\wt\nu-s\wt\mu)}+1}{e^{4\pi\wt\nu}-1}.
\ede
Apparently there is no $k$ dependence in this expression, and this is because we are considering the late-time limit where the particles are fully non-relativistic.
The real gauge bosons are produced when the physical momentum $|k\tau|=\mu$. After the production they are redshifted to all lower wave numbers. So we will restrict the above momentum integral within a sphere of radius $\mu$. Then we see that the gauge boson energy density is
\begin{align}
  \rh_A\simeq\int_0^\mu\FR{\di^3k}{(2\pi)^3}\sqrt{m^2+k^2} n_k =\FR{m_A^4}{16\pi^2}\Big[\varrho(1+2\varrho^2)\sqrt{1+\varrho^2}-\log\Big(\varrho+\sqrt{1+\varrho^2}\Big)\Big]e^{2\pi(\wt\mu-\wt\nu)}.
\end{align}
We note again that the late-time expansion (\ref{A_mode_late}) we used here is valid only when $|k\tau|\lesssim 1$ while we have performed the $k$ integral of the  resulting particle density $n_k$ up to $k=\mu$. We expect some corrections to this result when $\mu\gg H$ which is probably insignificant. Then we impose the constraint that $\rh_A$ is much smaller than the energy density during inflation,
\bge
\begin{aligned}
  &\rh_A\ll 3\Mp^2H^2.
\end{aligned}
\ede
One may consider a stronger condition that the gauge boson energy density $\rh_A$ should not affect the slow-roll potential of the inflaton. But this is unnecessary since we can always adjust the inflation potential a little bit so that the resulting power spectrum agrees with data even including the effects of gauge bosons.  

For $\varrho$ not too different from 1, namely $m_A\simeq \mu = \dot\phi_0/\Lambda$, we can see that the mass of the gauge boson can never be greater than $\dot\phi_0^{1/2}$ by the EFT constraint (\ref{eft_con}). So $m_A^4\lesssim \dot\phi_0^2=2\ep\Mp^2H^2\ll 3\Mp^2H^2$, where $\ep$ is the first slow-roll parameter. So the density $\rh_A$ is never larger than $3\Mp^2H^2$ without the exponential factor. Therefore it is only important to consider the parameters with $\wt\mu>\wt\nu$ or equivalently $\varrho>1$. In this case the density is roughly
\bge
\rh_A\simeq \FR{1}{16\pi^2}m_A^4\varrho^4e^{2\pi(\varrho-1)m_A/H}.
\ede
We also note that this bound becomes less constraining for lower scale inflation since we have shown that $\rh_A\lesssim  \varrho^4e^{2\pi(\varrho-1)m_A/H}\ep\rho$ by EFT bound. For lower scale inflation both $\rho$ and $\ep\propto \rho$ get smaller, and thus one could tolerate a larger exponential factor. 

\paragraph{Equilateral Non-Gaussianity.} As mentioned  in Sec.\;\ref{sec_est}, the same processes that contribute to the gauge boson signal also generate a ``background,'' which in the EFT limit contribute to the equilateral NG. Equilateral NG has been constrained by Planck measurement to be $f_\text{NL}^\text{(eq)}\lesssim \order{10}$. Therefore, using our estimate (\ref{f_b_bg_est}), we have the following constraint,
\begin{align}
\FR{1}{4}P_\zeta u^{-3} e^{6\pi(\varrho-1)m_A/H}<\order{10}.
\end{align}

\paragraph{Perturbativity.} The gauge boson signals were calculated using perturbation theory. At the diagram level, the gauge boson production manifests itself through the exponential enhancement of gauge boson propagators. A large exponential factor could invalidate the perturbative expansion in terms of diagrams. For example, we can consider the interaction vertex $g^2\si^2A^2$. Inspection of loop expansion with this vertex shows that the effective expansion parameter is 
\begin{align}
\frac{g^2H^2}{16 \pi^2 m_\si^2} e^{2\pi(\wt\mu-\wt\nu)}.
\end{align}
Pertubativity requires (in the limit $\mu, \ m_A \gg H$)
\begin{align}
\frac{g^2 H^2}{16 \pi^2m_\si^2} e^{2\pi(\wt\mu-\wt\nu)} < 1 \quad \to \quad \varrho < 1+ \frac{1}{\pi} \frac{H}{m_A} \log \frac{4 \pi m_\si}{g H}.
\end{align}
Of course this bound depends on the interactions. One could also consider other interactions such as $\phi F\wt F$ which is less important than the above one. We also note that the breakdown of perturbation expansion is not a physical problem. There could interesting effects in the strongly coupled region of parameter space which we leave for a future work.

\section{Discussions} 
\label{sec_discussions}

The main results of this paper are presented in Fig.~\ref{fig_fnl}  and \ref{fig_signal}. We see that in the parameter region of $\mu \sim m_A \gg H$, there is a window of opportunity in which the oscillatory signal of gauge boson production is observable by current or near future probes. $f_{\rm{NL}}^\text{(osc)}$ can be as large as $ {\mathcal{O}}(10)$. In addition, the gauge boson production will also produce NG in the equilateral limit. Parameter space with even larger oscillatory signal is already constrained by the current limit on the  equilateral NG. It is also noteworthy that the signal considered in this paper occupies the oscillation frequency region $4\lesssim \wt\nu\lesssim20$, with $f_\text{NL}^\text{(osc)}\gtrsim 0.1$, shown in Fig.~\ref{fig_signal}. This is a signal region distinct from models in the category of quasi-single-field-inflation, and those in which the inflaton has a coupling to the chiral current of massive fermions. 

In this paper, we have also developed a set of rules to estimate the size of the signal, with the exponential factor properly taken into account. This helped us to focus on a set of diagrams with dominant contribution. Through a more careful study of the late time behavior of the propagator and their time integral in App.\ \ref{app_mass_dep}, we also notice that the part of the propagator which is analytic in momentum can also contribute to the oscillatory signal, which has missed by previous studies. We also found that while the exponential factor can be reliably estimated, the power dependence on the mass parameters (gauge boson mass $m_A$ in our case) can not be captured by a simple counting. While the approximate calculation in Section~\ref{sec:explicit} improves upon the simple estimate, it is unlikely to contain the fully accurate power dependence on $m_A$. This underscores the important to perform a full calculation without approximation in order to set completely accurate constrains and make projections. This is a promising direction to pursue in the future. 

\paragraph{Acknowledgment.} We thank Xingang Chen, Junwu Huang, Matt Reece, and Yi Wang for useful discussions. LTW is supported by the DOE grant DE-SC0013642. 

\begin{appendix}

\section{More on Estimating the Signal Strength}
\label{app_mass_dep}

In \cite{Wang:2019gbi} we showed how to estimate the signal strength by assigning simple factors to each vertex and propagator in the SK diagram. This simple shortcut can get correctly the exponential dependence from  propagators as well as the power dependence from vertices. However, it is generally difficult to get the power dependence from the propagator correctly. In this appendix we elaborate on this issue. The main conclusions are: 
\begin{enumerate}
  \item It is not sufficient to include power dependence on the mass parameters from the propagators and the vertices, since the SK time integrals can introduce additional power dependence (although these integrals do not introduce additional exponential dependence.).  It seems there is no simple way to count the powers from the time integrals. 
    
  \item For heavy propagator with large mass $m\gg H$ that can be approximated as EFT local operators, one can show that the power dependence on the mass is always $1/m^2$, by either EFT argument or explicit calculation. There could also be exponential dependence for EFT part due to chemical potential which can also be reliably estimated. 
    
  \item For non-local ``on-shell'' propagators, there is no simple way to estimate the power dependence correctly. It is likely that even more careful calculation with late-time expansion of propagators cannot get the powers right, either. However, we expect the late-time expansion can at least capture a fraction of oscillation signals, so it is still a useful way to evaluate the signal strength. 
\end{enumerate}

To illustrate these point we will first consider a simpler case in flat space, and then go to the inflation background.

\paragraph{Mass dependence in flat-space correlators.} First we will show that the time integral can generate additional power dependence. For this purpose it is helpful first to look at the flat-space example. In flat space there is no oscillation features in the correlation function, and thus we will only consider the ``local'' part.

In flat space the propagator of a scalar field $\si$ of mass $m_\si$ can be written as
\bge
  D_>(k;t_1,t_2)=\FR{1}{2E_k}e^{-\ii E_k(t_1-t_2)},
\ede
where $E_k=\sqrt{m_\si^2+k^2}$. In the large $m$ limit the propagator goes like $1/m$, the same as the propagator in inflation. So naively we would estimate the contribution of such a propagator as $1/m$ for large $m$. However, we know that the correct EFT limit is $1/m^2$. So where is the additional power dependence from?\footnote{In usual treatment in 4-momentum space, the propagator is $1/(k^2-m^2)$ and the EFT limit $1/m^2$ is manifest. But we are now working in the 3-momentum space where the time direction is not Fourier transformed.} 

The answer is that the time integral will contribute another power. To see this, we calculate the correlator of 4 light scalar fields $\phi$ at $t=0$ connected by an $s$-channel $\si$, with $m_\phi\ll m_\si$ and vertex $\lam\phi^2\si$. Using the diagrammatic rule, the correlator is
\begin{align}
  \mathcal{I}=&~(\ii\lam)^2\sum_{\mathsf{a},\mathsf{b}=\pm 1}\mathsf{a b}\int_{-\infty}^0\di t_1\di t_2 \FR{1}{2E_1}\FR{1}{2E_2}\FR{1}{2E_3}\FR{1}{2E_4} e^{\ii a (E_1+E_2)t_1+\ii b(E_3+E_4)t_2}D_\mathsf{ab}(k_s;t_1,t_2).
\end{align}
Here $E_i=\sqrt{k_i^2+m_\phi^2}~(i=1,\cdots, 4)$ and $E_s=\sqrt{k_s^2+m_\si^2}$. Here we can already see that the time integral will introduce additional power dependence on $m_\si$ at large $m_\si$ such as $\int \di t_1 e^{\ii E_s t_1}\sim E_s^{-1}\sim m_\si^{-1}$. However, superficially we would expect two powers of $m_\si^{-1}$ being introduced since we have two time integrals. This combined with $m_\si^{-1}$ would give $1/m_\si^3$ behavior, not in agreement with EFT counting $1/m_\si^2$. But if we look at the integral more closely, by writing $\mathcal{I}=\lam^2(\mathcal{I}_T+\mathcal{I}_N)$ in terms of the time-ordered part $\mathcal{I}_T$ and non-time-ordered part, $\mathcal{I}_N$, we will get,
\begin{align}
  \mathcal{I}_T=&-\FR{1}{2E_1\cdots 2E_4}\FR{1}{2E_s}\,2\text{Re}\,\bigg[\int_{-\infty}^0\di t_1\int_{-\infty}^{t_1}\di t_2\,e^{\ii(E_{12}-E_s)t_1+\ii(E_s+E_{34})t_2}\n\\
  &~+\int_{-\infty}^0\di t_2\int_{-\infty}^{t_2}\di t_1\,e^{\ii (E_{12}+E_s)t_1+\ii (E_{34}-E_s)t_2}\bigg]\n\\
  =&~\FR{1}{2E_1\cdots 2E_4}\FR{1}{E_s(E_{12}+E_{34})}\bigg[\FR{1}{E_s+E_{12}}+\FR{1}{E_s+E_{34}}\bigg],\\
  \mathcal{I}_N=&~\FR{1}{2E_1\cdots 2E_4}\FR{1}{2E_s}\,2\text{Re}\,\int_{-\infty}^0\di t_1\di t_2\,e^{\ii (E_{12}+E_s)t_1-\ii (E_{34}+E_s)t_2}\n\\
  =&~\FR{1}{2E_1\cdots 2E_4}\FR{1}{E_s(E_{12}+E_s)(E_{34}+E_s)},
\end{align}
where $E_{12}\equiv E_1+E_2$ and $E_{34}=E_3+E_4$. So it is the time-ordered part $\mathcal{I}_T\sim 1/m_\si^2$ that gives the correct EFT behavior for large $m_\si$, while the non-time-ordered part $\mathcal{I}_N\sim 1/m_\si^{3}$ gives subleading contribution. Our previous naive guess implicitly assumed that the two time integrals can be factorized and thus applies only to $\mathcal{I}_N$.

\paragraph{Mass dependence in inflationary correlators.} The above example shows that the time-ordered integral is important to get the correct power dependence. Similar calculation in the inflation background can also be done, giving similar result. As a demonstration, we use the example of  quasi-single-field inflation with one massive field $\sigma$.  The relevant operator is $(\partial \phi)^2 \sigma$. Expanding around background value, $\phi = \phi_0 + \delta \phi$ and $\sigma = \sigma_0 + \delta \sigma$, it gives rise to a 3-point vertex $(\delta \phi')^2 \delta \sigma$ and a two point mixing $\delta \phi' \delta \sigma$. Putting these together, we can form a 3-point diagram with one massive propagator $\sigma$ in the middle. From naive estimate in the large mass limit $\wt\nu = (m_\sigma^2/H^2 - 9/4)^{1/2} \gg 1$, the leading EFT piece scales like $\wt\nu^{-2}$, while the signal scales as $e^{-\pi\wt\nu}$.

Analogous to the flat space example,  we will separate the integral into a time-ordered piece $\mathcal{I}_T$ and non-time-ordered piece $\mathcal{I}_N$.
\begin{align}
  \mathcal{I}_T=&~\FR{H^6}{8k_1k_2k_3}2\,\text{Re}\bigg[\int_{-\infty}^0\FR{\di\tau_1}{|H\tau_1|^2}\int_{-\infty}^{\tau_1}\FR{\di\tau_2}{|H\tau_2|^3}\,\tau_1^2\tau_2e^{\ii(k_1+k_2)\tau_1+\ii k_3\tau_2}D_>(k_3;\tau_1,\tau_2) \n\\
  &+\int_{-\infty}^0\FR{\di\tau_1}{|H\tau_1|^2}\int_{\tau_1}^0\FR{\di\tau_2}{|H\tau_2|^3}\,\tau_1^2\tau_2e^{\ii(k_1+k_2)\tau_1+\ii k_3\tau_2}D_>^*(k_3;\tau_1,\tau_2)\bigg].
\end{align}
\begin{align}
  \mathcal{I}_N=&~\FR{H^6}{8k_1k_2k_3}2\,\text{Re}\bigg[\int_{-\infty}^0\FR{\di\tau_1}{|H\tau_1|^2}\int_{-\infty}^{0}\FR{\di\tau_2}{|H\tau_2|^3}\,\tau_1^2\tau_2e^{\ii(k_1+k_2)\tau_1-\ii k_3\tau_2}D_>^*(k_3;\tau_1,\tau_2)\bigg].
\end{align}
We have treated the inflaton as massless. $D_>(k;\tau_1,\tau_2)\equiv u(\tau_1,k)u^*(\tau_2,k)$ is the propagator of the massive scalar field, and $u$ is its mode function, given by
\begin{align}
\label{massivescalarmode}
  u(\tau,\mb k)=\FR{\sqrt{\pi}}{2}e^{\ii\pi(\nu/2+1/4)}H(-\tau)^{3/2}\text{H}_\nu^{(1)}(-k\tau).
\end{align}

The mode function $u(\tau,k)$ in this case contains a Hankel function which makes a direct integration difficult. To make progress, we expand it in the late-time limit. This is not entirely valid, and we will show how far we can get.
\begin{align}
  u(\tau, k)\sim \FR{H}{2\sqrt{\pi}}(-\tau)^{3/2}\Bigg[\Gamma(-\ii\wt\nu)e^{\pi\wt\nu/2}\Big(\FR{-k\tau}{2}\Big)^{\ii\wt\nu}+\Gamma(\ii\wt\nu)e^{-\pi\wt\nu/2}\Big(\FR{-k\tau}{2}\Big)^{-\ii\wt\nu}\Bigg].
\end{align}
From this we can again construct a local propagator and a non-local propagator,
\begin{align}
  D_>^{(\text{local})}(k,\tau_1,\tau_2)=&~\FR{H^2}{4\pi}(\tau_1\tau_2)^{3/2}\Gamma(-\ii\wt\nu)\Gamma(\ii\wt\nu)e^{\pi\wt\nu}\Big(\FR{\tau_1}{\tau_2}\Big)^{\ii\wt\nu},\n\\
  D_>^{(\text{nonlocal})}(k,\tau_1,\tau_2)=&~\FR{H^2}{4\pi}(\tau_1\tau_2)^{3/2}\bigg[\Gamma^2(-\ii\wt\nu)\Big(\FR{k^2\tau_1\tau_2}{4}\Big)^{\ii\wt\nu}+\Gamma^{2}(+\ii\wt\nu)\Big(\FR{k^2\tau_1\tau_2}{4}\Big)^{-\ii\wt\nu}\bigg],
\end{align}
where we have neglected a term in $D_>^\text{(local)}$ that is further suppressed by a factor of $e^{-2\pi\wt\nu}$.

The contribution to the time ordered integral  $ \mathcal{I}_T$ from the local propagator is
\begin{align}
\label{eq:local}
  \mathcal{I}_T^\text{(local)}=\FR{1}{64H^3k_1^4k_3^2}\bigg\{\FR{1}{\wt\nu^2}\bigg(\FR{k_3}{k_1}\bigg)-\pi \wt\nu e^{-\pi\wt\nu}\text{Im}\,\bigg(\FR{k_3}{2k_1}\bigg)^{1/2-\ii\wt\nu} \bigg\},
\end{align}
while the non-local propagator gives
\begin{align}
\label{eq:nonlocal}
  \mathcal{I}_{T}^\text{(nonlocal)}=-\FR{1}{64H^3k_1^4k_3^2}\cdot 4\pi\wt\nu e^{-\pi\wt\nu}\text{Re}\bigg(\FR{k_3}{2k_1}\bigg)^{1/2+\ii\wt\nu}.
\end{align}
The non-time-ordered integral always gives more suppressed result. A few comments are in order.
\begin{enumerate}
\item The first term in (\ref{eq:local}), $\propto m_\sigma^{-2}$ in the large mass limit, reproduce the EFT result for massive propagator. 
\item The result should be compared with the full result in \cite{Chen:2015lza}, in which the full propagator was used without making late-time expansion. Then in the large mass limit $\wt\nu\gg 1$, the signal scale as $\wt\nu^{3/2}e^{-\pi\wt\nu}$ instead of the late-time result $\wt\nu e^{-\pi\wt\nu}$. The mismatch of the powers signifies a failure of the late-time expansion.  
\item The second term in  (\ref{eq:local}) shows that the local part of the propagator (which is analytic in $k$) can also contribute to the oscillatory signal after time integral. A similar result was also observed in \cite{Chen:2015lza} for the more suppressed signal in the non-time-ordered integral.

\end{enumerate}

\end{appendix}

\providecommand{\href}[2]{#2}\begingroup\raggedright\endgroup

\end{document}